\pdfoutput=1
\documentclass[%
 reprint,
superscriptaddress,
 amsmath,amssymb,
 aps,
prx,
floatfix,
]{revtex4-2}
\usepackage{graphicx}
\usepackage{dcolumn}
\usepackage{bm}
\usepackage{amsthm}
\usepackage{commath}
\usepackage{mathtools}
\usepackage{graphicx}
\usepackage[colorlinks=true, hyperindex, breaklinks, linkcolor=blue, urlcolor=blue, citecolor=blue]{hyperref} 
\usepackage{amsmath,amsfonts,amssymb,color,graphicx,xcolor,physics}
\usepackage{tikz}
\usepackage{float}
\makeatletter
\let\newfloat\newfloat@ltx
\makeatother
\usepackage{algorithm} 
\usepackage{algorithmicx}
\usepackage[noend]{algpseudocode} 
\usepackage[normalem]{ulem}

\newcommand{\kn}[1]{\textcolor{blue} {#1}}

\newcommand{\rev}[1]{\textcolor{black} {#1}}

\newcommand{\dm}[1]{\ket{#1}\bra{#1}}
\newcommand{\st}[1]{\ket{#1}}
\newcommand{\sbkt}[1]{\left(#1\right)}
\newcommand{\mbkt}[1]{\left[#1\right]}
\newcommand{\sandwich}[3]{\bra{#1}#2 \ket{#3}}
\newcommand{\bigO}[1]{\mathcal{O}\left(#1\right)}

\newcommand\nonpfrate[1]{\gamma_{X, Y}}
\makeatletter
\newcommand*{\rom}[1]{\expandafter\@slowromancap\romannumeral #1@}
\makeatother

\def\algbackskip{\hskip-\ALG@thistlm}

\begin{document}

\preprint{APS/123-QED}

\title{Cross-resonance control of an oscillator with an auxiliary fluxonium qubit}

\author{Guo Zheng} 
\affiliation{Pritzker School of Molecular Engineering, The University of Chicago, Chicago 60637, USA}
\affiliation{AWS Center for Quantum Computing, Pasadena, CA 91125, USA}

\author{Simon Lieu}
\affiliation{AWS Center for Quantum Computing, Pasadena, CA 91125, USA}

\author{Emma L.~Rosenfeld}
\affiliation{AWS Center for Quantum Computing, Pasadena, CA 91125, USA}

\author{Kyungjoo Noh}
\email{nkyungjo@amazon.com}
\affiliation{AWS Center for Quantum Computing, Pasadena, CA 91125, USA}

\author{Connor T.~Hann}
\affiliation{AWS Center for Quantum Computing, Pasadena, CA 91125, USA}

\date{\today}

\begin{abstract}
The conditional displacement (CD) gate between an oscillator and a discrete-variable auxiliary qubit plays a key role in quantum information processing tasks, such as enabling universal control of the oscillator and longitudinal readout of the qubit. However, the gate is unprotected against the propagation of auxiliary qubit decay errors and hence not fault-tolerant. Here, we propose a CD gate scheme with fluxonium as the auxiliary qubit, which has been experimentally demonstrated to have a large noise bias and millisecond-level lifetimes. The proposed gate is applied cross-resonantly by modulating the external flux of the fluxonium at the frequency of the target oscillator, which requires minimal hardware overhead and does not increase sensitivity to decoherence mechanisms like dephasing. We further provide a perturbative description of the gate mechanism and identify the error budget. Additionally, we develop an approximate procedure for choosing device and gate parameters that optimizes gate performance. Following the procedure for multiple sets of fluxonium parameters from the literature, we numerically demonstrate CD gates with unitary fidelity exceeding 99.9\% and gate times of hundreds of nanoseconds.

\end{abstract}

\maketitle


\section{Introduction}

Bosonic modes, also known as quantum harmonic oscillators, are alternative solutions for storing and manipulating quantum information. Experimentally, bosonic modes can be realized in superconducting~\cite{PhysRevB.94.014506, PhysRevLett.107.240501, PhysRevB.86.100506, PhysRevApplied.13.034032, PRXQuantum.4.030336} and mechanical architectures~\cite{Arrangoiz_Arriola_2019, Jost_2009}, and many of these oscillators have exhibited long coherence times on the order of milliseconds. Moreover, a major advantage of bosonic modes compared to conventional qudits is their infinite-dimensional Hilbert spaces, which enables an additional layer of quantum error correction, known as bosonic error correction~\cite{Mirrahimi_2014, Puri_2017, PhysRevX.6.031006, Gottesman_2001, Schlegel_2022}. Bosonic error correction is capable of further suppressing the physical error rates to significantly reduce the resource overheads required for fault-tolerant quantum computations~\cite{PRXQuantum.3.010329, PhysRevX.9.041053, PhysRevA.101.012316, PRXQuantum.3.010315, Xu_2023}. In recent years, significant experimental progress has been made towards the realization of bosonic codes and surpassing the break-even point~\cite{Grimm_2020, deNeeve_2022, Ofek_2016, Hu_2019, Campagne_Ibarcq_2020, Lescanne_2020, Fl_hmann_2019, Sivak_Nature_2023}. 

To process quantum information, it is critical to have sufficient control of individual components of a quantum hardware device. In particular, since an oscillator is a linear system, it is paramount to include nonlinear elements for universal control. A common solution is to couple the oscillator to a nonlinear system, such as a qubit, and control it through some entangling dynamics. Among all entangling operations between a qubit and an oscillator, the conditional displacement (CD) gate plays a unique role in many information processing schemes. For example, together with qubit rotations, CD gates enable universal control of the oscillator~\cite{Eickbusch_2022_NatPhys, diringer2023conditional}. It also serves as a key ingredient in the stabilization of the Gottesman-Kitaev-Preskill (GKP) codes~\cite{Royer_2020, Campagne_Ibarcq_2020, Hastrup_2021} and in the state preparation of the binomial codes and the cat codes~\cite{Eickbusch_2022_NatPhys}. In addition, when the information is encoded in the qubit, the CD gate can be applied for qubit state readout~\cite{PhysRevLett.115.203601, PhysRevLett.122.080502}. \rev{Experimentally, the CD gate has been demonstrated in, for example, circuit QED platforms~\cite{Hacohen2016Nature, PhysRevLett.120.040505, PhysRevLett.122.080502}.} 

However, state-of-the-art implementations of the CD gate have major practical limitations. The commonly-adopted gate design~\cite{Eickbusch_2022_NatPhys,Campagne_Ibarcq_2020} achieves a fast, so-called ``echoed-CD gate" between a high quality-factor cavity and a transmon qubit, with gate speed not limited by the dispersive coupling strength. As a tradeoff, the oscillator needs to be displaced to a large amplitude, and stay in highly excited states during the gate. This can then degrade the gate fidelity since the impacts of some error sources, such as dephasing and Kerr nonlinearities, are amplified as average photon number increases. While the dephasing of cavities is in general much smaller than loss, it is known that certain codes like the GKP codes are more sensitive to dephasing than loss~\cite{PRXQuantum.2.020101}. Therefore, it is desirable to operate the oscillator with as low photon numbers as possible while implementing a CD gate. Another general shortcoming of schemes based on qubit control is that they are not fault-tolerant against auxiliary qubit decay. Often times, since the oscillators, such as three-dimensional electromagnetic cavities, have physical error rates significantly smaller than that of the auxiliary qubits, such as transmons, the performance of the bosonic code is severely limited by the coherence times of the auxiliary qubit~\cite{Royer_2020}. While there exist methods, such as $\chi$-matching~\cite{Rosenblum_Science_2018, Reinhold_NaturePhysics_2020}, to protect the oscillator against auxiliary qubit decay, the consequent engineering challenges and overhead from the increased control complexity remain daunting.

To address these limitations, in this work we propose and analyze a cross-resonance (CR) implementation of a CD gate between a fluxonium qubit and an oscillator. The CR gate between qubits~\cite{Paraoanu_2006, PhysRevA.100.012301, PhysRevB.81.134507, PhysRevA.101.052308, PhysRevA.102.042605, Chow_2011} refers to a gate activated by driving the control qubit at the frequency of the target qubit, and it has been a promising mechanism for realizing multi-qubit operations in transmon-based architectures~\cite{PhysRevLett.129.060501, PhysRevA.93.060302, Jurcevic_2021}. It is popular because it only requires microwave control, with minimal hardware overhead. Here, by generalizing CR gates to control an oscillator, we propose a CD gate which avoids occupying high-energy states of the oscillator. Our CD gate applies a microwave drive at the frequency of the oscillator to each element. The drives introduce a displacement only when the auxiliary qubit is in the excited state, by bringing a sideband of the coupled system into resonance. \rev{A similar CR implementation of the CD gate was experimentally demonstrated in Ref.~\cite{PhysRevLett.122.080502} in the context of transmon readout.}

Another key ingredient to our scheme is the fluxonium qubit~\cite{Manucharyan_2009}, which has received considerable attention as a promising superconducting-qubit candidate. Recently, there has been substantial theoretical and experimental progress in controlling fluxonium, including single and two-qubit gates~\cite{Zhang_2021, PhysRevLett.129.010502, PRXQuantum.3.037001, PhysRevApplied.20.024011, PhysRevApplied.20.024011, PhysRevX.13.031035, dogan2022demonstration, PhysRevApplied.18.034063} and its applications in hybrid architectures~\cite{PhysRevResearch.4.043127, PRXQuantum.4.040342}. The fluxonium circuit consists of a Josephson junction, a capacitor, and a superinductor connected in parallel, with flux tunability of the spectrum. When the external flux equals half-integer flux quantum, also known as the ``sweet spot", the fluxonium qubit enjoys a suppression of decoherence from flux noise~\cite{PRXQuantum.3.037001, PhysRevLett.129.010502, PhysRevX.9.041041}. For example, at the sweet spot, fluxonium coherence times have been demonstrated to be on the order of milliseconds~\cite{PhysRevLett.130.267001, PhysRevX.9.041041}. Given its large anharmonicity, it is also more robust to population leakage during gates compared to the transmon. Remarkably, when the fluxonium operates off sweet spot, it has a biased error channel such that the $T_1$ time can be significantly extended at the cost of a reduced coherence time $T_2^*$~\cite{PhysRevLett.120.150503, PhysRevX.9.041041}. \rev{This property of fluxonium can be exploited to provide significant benefit in certain applications. One example is when it is adopted as the auxiliary qubit of the CD gate, whose performance is much more sensitive to auxiliary qubit energy decay than dephasing.} Similar ideas of \rev{an auxiliary qubit with suppressed bit-flip rate} have been explored in the context of, for example, cat codes~\cite{PhysRevX.9.041009, ding2024quantumcontroloscillatorkerrcat}, but fluxonium in particular can be an economic choice that circumvents additional circuit complexity. 

We analyze the performance of our proposed CD gate both analytically and numerically, identifying dominant coherent error mechanisms and developing an approximate procedure for optimizing device and control parameters. To suppress errors caused by dispersive shifts and to stay in a low photon space, we incorporate techniques such as echoing~\cite{PhysRevA.87.030301, Jurcevic_2021, Eickbusch_2022_NatPhys} and selective darkening (SD)~\cite{de_Groot_2010, de_Groot_2012}. We further analyze the sources of coherent errors over a wide range of gate durations, by separately considering on and off resonance terms in the drive Hamiltonian. Our error analysis motivates a comprehensive procedure for choosing oscillator frequency, coupling strength, and drive strength subject to a given fluxonium parameter set. While the procedure does not guarantee optimality, it provides a systematic approach for determining relevant operational parameters with a reasonable gate fidelity, which otherwise would require a computationally intensive search through a large parameter space. As some examples, we demonstrate such a capability by examining different sets of fluxonium parameters, both on and off the sweet spot, that have appeared in literature. We show for all examples that the CD gates can be realized in a few hundreds of nanoseconds with unitary infidelity lower than $10^{-3}$. 

\rev{One of the key contributions of our work lies in the analysis of the gate mechanism. Most cross-resonance schemes focus on either transmons~\cite{PhysRevLett.122.080502, Paraoanu_2006, PhysRevA.100.012301, PhysRevB.81.134507, PhysRevA.101.052308, PhysRevA.102.042605, Chow_2011} or using only low excited states of the fluxonium~\cite{dogan2022demonstration, PhysRevApplied.20.024011, PhysRevResearch.4.043127}. Compared to the transmon, the strong nonlinearity in the fluxonium Hamiltonian makes it highly challenging to obtain accurate predictions from perturbative analysis on the Hamiltonian level. In contrast, our analysis stems from first principles~\cite{PhysRevA.102.042605} and relies upon branch analysis~\cite{PhysRevX.14.041023, Shillito_2022, Boissonneault_2010} of the complicated structure of the hybridized system. This non-perturbative analysis methodology reveals that cross resonance control can in principle be realized with the assistance of any excited states of the fluxonium, and we demonstrate high-performance conditional displacements relying on higher excited states.}

This paper is organized as follows. In Sec.~\ref{sec:model_gate}, we introduce our system Hamiltonian, drive scheme, and techniques such as echoing and selective darkening. Sec.~\ref{sec:params_choice} provides intuition on our choice of operational parameters. In particular, we go through the parameters and determine them individually, in order of increasing tunability for a typical experiment. In Sec.~\ref{sec:error_budget}, we analyze various sources of the gate infidelity by isolating the effects of different classes of undesired Hamiltonian terms. This analysis in turn justifies the parameter selection procedure we outline. We apply this approach to a few other sets of fluxonium parameters and demonstrate the gate performance in Sec.~\ref{sec:survey}. Finally, Sec.~\ref{sec:conclusion} summarizes this work.

\section{Model and gate scheme\label{sec:model_gate}}

\begin{figure}[t!]
    \centering
    \includegraphics[width = 0.45 \textwidth]{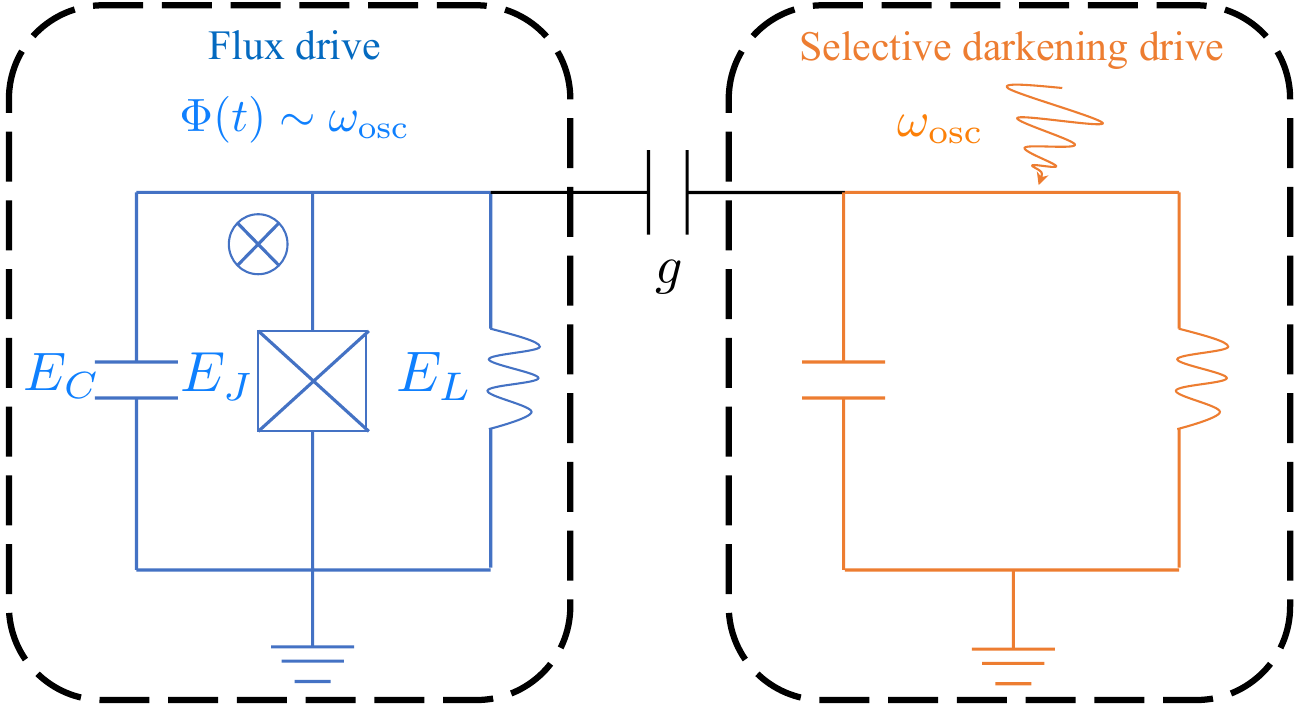}
     \caption{The circuit diagram of the coupled system. The left (right) box labels the fluxonium (oscillator). The external flux of the fluxonium is modulated at the oscillator's frequency, $\omega_{\text{osc}}$. A direct drive, $\hat{a} + \hat{a}^\dagger$, is applied to the oscillator at the same frequency for selective darkening. }
     \label{fig:circuit_diagram}
\end{figure}

Cross-resonance gates apply pulses to the control system at the frequency of the desired dynamics in the target system. At a high level, they utilize the control system as a quantum switch, such that the desired interaction is only activated when the control system is in the designated logical state. The gate speeds are direct results of the hybridization between the control and target systems, which in turn depends on the states' energy gaps and matrix elements of the coupling Hamiltonian. By carefully designing the system parameters, we can selectively induce significant hybridization involving only one of the computational states, thereby realizing efficient conditional dynamics. To be specific, in the following sections, we assume a fluxonium qubit for auxiliary control of an oscillator as the target.

\subsection{System Hamiltonian}
We consider a fluxonium qubit capacitively coupled to an oscillator. The full circuit Hamiltonian can be written as 
\begin{eqnarray}\label{eq:full_Ham}
    \hat{H}/\hbar &=& \hat{H}_f/\hbar + \omega_{\text{osc}} \hat{a}^\dagger \hat{a} + ig\hat{n}\left(\hat{a} - \hat{a}^\dagger\right) + \hat{H}_{\text{d}}/\hbar,
\end{eqnarray}
where $\omega_{\text{osc}}$ is the oscillator frequency and $g$ is the coupling strength (see the footnote in \footnote{ Note that in some conventions the $g$ coupling strength is defined as the energy coefficient of the $(\hat{b} - \hat{b}^\dagger)$ term  that arises when expressing the charge operator in terms of creation and annihilation operators of an anharmonic oscillator: $\hat{n} = in_0\left(\hat{b} - \hat{b}^\dagger\right)$. This differs by the zero-point energy term $n_0 := \left(E_L / (32E_c)\right)^{1/4}$. The zero-point fluctuation caused by impedance of the linear oscillator mode $\hat{a}$ is absorbed in $g$.}). The first three terms represent the static system Hamiltonian comprising of the fluxonium, the oscillator, and their capacitive coupling, and the last term corresponds to the drive Hamiltonian. The fluxonium Hamiltonian is
\begin{eqnarray}
    \hat{H}_f &=& 4E_C \hat{n}^2 + E_J \cos{\hat{\varphi} } + \frac{E_L}{2}\left(\hat{\varphi} - 2\pi \frac{\Delta \Phi}{\Phi_0}\right)^2,
\end{eqnarray}
where $E_{C}, E_J, E_L$ are the fluxonium's charging, Josephson, and inductive energies respectively. Here, $\Phi_0 = h/2e$ is the superconducting flux quantum, and $\Delta \Phi$ represents the static deviation of the external flux from the half flux quantum ($\Phi_{0}/2$) sweet spot. The external flux deviation is placed in the last term to properly model typical fluxonium experiments with time-dependent flux~\cite{PhysRevB.99.174512, PhysRevApplied.19.034031}. A circuit diagram illustrating the individual systems, the capacitive coupling, and the drive tones is given in Fig.~\ref{fig:circuit_diagram}.

The major difference between fluxonium qubits and conventional transmon qubits is that the potential well in the phase basis is now a combination of a parabolic potential and a cosine potential. The half-flux sweet spot corresponds exactly to the scenario where a maxima of the cosine potential coincides with the minima of the parabolic potential. In the heavy fluxonium regime, i.e. when $E_J/E_C\gg 1$, the computational states are the even and odd superposition of states localized in each well. It has been shown that fluxonium at sweet spot is first-order insensitive to decoherence mechanisms such as flux noise~\cite{PhysRevX.9.041041}. Moreover, since the frequency of heavy fluxonium at sweet spot tends to be low (typically much less than $1$ GHz) and the charge matrix elements are reduced, the dielectric loss is also suppressed~\cite{PhysRevX.9.041041, PhysRevLett.120.150504}. When the fluxonium is operated off sweet spot, the symmetry-protected robustness to flux noise is lost, which results in decreased coherence times. Also, because of the broken symmetry, the computational states become well-localized in the left and right wells, the so-called ``fluxon" excitations. Adjusting the flux away from the sweet spot further reduces the charge matrix element of the computational transition, by suppressing fluxon exchange due to the broken degeneracy between the left and right wells. This further protects the qubit from dielectric loss, increasing $T_1$~\cite{PhysRevX.9.041041, PhysRevLett.120.150504}. Intuitively, for localized states in the cosine minima separated by an energy gap, a local perturbation is unlikely to cause energy decay. The asymmetry between the increased dephasing noise rates and decreased energy dissipation bears similarity to the noise bias in cat codes~\cite{Mirrahimi_2014, PhysRevX.9.041009}.
The biased-noise property can be exploited when the target task has different sensitivities to each type of noise~\cite{PhysRevX.9.041009}. One example is the CD gate, where an auxiliary qubit decay event leads straight to a large deviation between the target and actual displacement while auxiliary qubit dephasing is less destructive~\cite{Royer_2020}. For a more detailed analyses of the noise properties of fluxoniums, we refer readers to Refs.~\cite{PhysRevX.9.041041, PhysRevApplied.20.034016, randeria2024dephasing, Hassani_2023,Masluk_thesis, Vladimir_Science_2009} and references therein.


The static coupled system Hamiltonian can be diagonalized as
\begin{eqnarray}
    \hat{H}_{\text{sys}}/\hbar&:=&\hat{H}_f/\hbar + \omega_{\text{osc}} \hat{a}^\dagger \hat{a} + ig\left(\hat{a} - \hat{a}^\dagger\right)\hat{n}\\
    &=& \sum_{i, n} E_{i, n} \ket{\phi_{i, n}}\bra{\phi_{i, n}},
\end{eqnarray}
where the eigenbasis are labelled by the dominant product state in the bare basis under the assumption of weak hybridization. Throughout the paper, we apply consistent state labeling such that the first and second index represents the fluxonium and oscillator states, respectively. The eigenbasis $\set{\ket{\phi_{i, n}}}$ and the bare basis $\set{\ket{i}\otimes \ket{n}}$ are related through a unitary transformation given by $\hat{U}_0$ such that 
\begin{eqnarray}
    \ket{\phi_{i, n}} = \hat{U}_0 \ket{i} \otimes \ket{n}.
\end{eqnarray}
Equivalently, for any operator $\hat{O}$, the matrix elements are equivalently transformed; $\bra{\phi_{i, n}}\hat{U}_0 \hat{O} \hat{U}_0^\dagger \ket{\phi_{i, n}} =  \bra{i} \otimes \bra{n}\hat{O}\ket{i} \otimes \ket{n}$. Experimentally, drives are applied in the bare basis, while states are prepared and measured in the eigenbasis. In this work, we are interested in the CD gate or, more precisely, the Echoed-CD (ECD) gate~\cite{Eickbusch_2022_NatPhys} in the eigenbasis, $\hat{U}_0 \hat{\text{ECD}}_\alpha \hat{U}_0^\dagger$, where 
\begin{eqnarray}\label{eq:original_CD}
    \hat{\text{ECD}}_\alpha := \vert 1\rangle \langle 0 \vert\otimes \hat{D}_{\frac{\alpha}{2}} + \vert 0\rangle \langle 1 \vert\otimes \hat{D}_{-\frac{\alpha}{2}}
\end{eqnarray}
with $\hat{D}_\beta := e^{\beta \hat{a}^\dagger - \beta^\ast \hat{a}}$ as the linear displacement operator. The ECD gate refers to the conventional CD gate with \rev{echoing}, i.e.~a Pauli-X is applied to the auxiliary qubit at the middle of the gate duration. Echoing mitigates errors like oscillator dephasing caused by dispersive shifts. The bare basis CD gate can be generated by a Hamiltonian of the form $\hat{\sigma}_z \left(\hat{a}+\hat{a}^\dagger \right)$, where $\hat{\sigma}_z$ is the Pauli-Z operator in fluxonium's computational space. In general, if the displacement rates of the oscillator differ for the fluxoniums' $\ket{0}$ and $\ket{1}$ states, the resulting unitary is equivalent to a CD gate modulo an unconditional, linear displacement on the oscillator, which can easily be cancelled out by a separate linear drive to recover the $\text{CD}$ operator.

\begin{figure}[t!]
    \centering
    \includegraphics[width = 0.45 \textwidth]{ 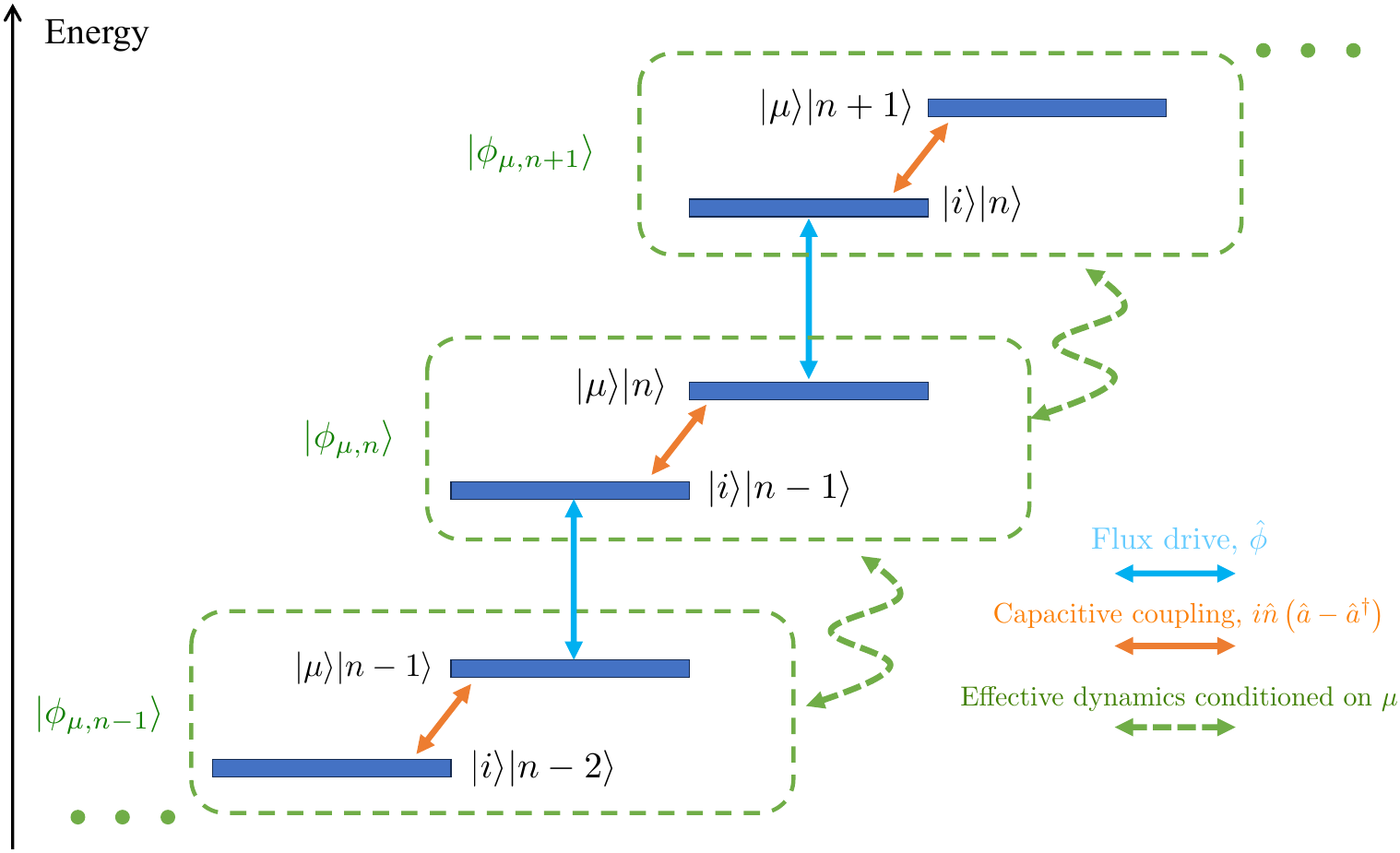}
     \caption{Level diagram explanation of gate mechanism. Here, we assumed the gate is activated by a pair of fluxonium states, $\sbkt{\mu, i}$, and, without loss of generality, $\Delta_{\mu, i}\geq \omega_{\text{osc}}$. For product states, the first and second state label the states of the fluxonium and the oscillator respectively. Here, only the relevant product state components of the eigenstates are labelled. In each green box, only the two main ingredients that give rise to the gate are shown. The blue and orange double arrows indicate the flux drive and the capacitive couplings respectively. }
    \label{fig:gate_int}
\end{figure}

\subsection{Gate scheme}

The CD gate is achieved through driving the system with
\rev{
\begin{eqnarray}
    \hat{H}_{\text{d}}/\hbar = f(t)\hat{\varphi} + f^\prime(t)\left(\hat{a} + \hat{a}^\dagger\right)\label{eq:drive_Hamiltonian}
\end{eqnarray}}
\rev{
where both the fluxonium and the oscillator are driven. In particular, as shown in the first term, the fluxonium is driven} through a flux drive, which can be applied by modulating the external flux, $\Delta \Phi$. See Appendix~\ref{sec:flux_drive_der} for a more detailed derivation with the time-dependent flux placed at the proper location~\cite{PhysRevApplied.19.034031}. \rev{The direct drive on the oscillator, shown in the second term, is applied to perform a technique called selective darkening (SD)~\cite{de_Groot_2010, de_Groot_2012, PhysRevApplied.18.034063}. As described in detail later, the direct drive should have the same time dependence as the flux drive. Without loss of generality, we set
\begin{eqnarray}
    f(t) = A \Omega(t)\cos\omega_{d}t.
\end{eqnarray}
Here, $A$ is the drive amplitude, $\Omega(t)$ is a pulse envelope with unit maximum magnitude. SD then requires $f^\prime(t) = c_{\text{SD}}f(t)$, where the coefficient $c_{\text{SD}}$ is determined below.}

If we disregard the direct drive on the oscillator for now and focus on the flux drive, the gate mechanism is illustrated schematically through Fig.~\ref{fig:gate_int}. The CD gate can be understood as an effective dynamics induced by the hybridization of a pair of fluxonium states, $\ket{\mu}$ and $\ket{i}$, with the former being one of the computational states and the latter outside the computational subspace. Through the hybridization introduced by the coupling term $\propto g$, the $\varphi$ operator in the eigenbasis contains matrix elements $\langle \hat{\varphi}  \rangle_{\mu, n}\propto\ket{\phi_{\mu, n-1}}\bra{\phi_{\mu, n}} + \text{h.c.}$, with coefficients depending on $\mu$, i.e., fluxonium-conditional displacement of the oscillator (see green arrows, Fig.~\ref{fig:gate_int}). In the following, we label the pair of level indices, $(\mu, i)$, to be the transition that induces the gate. The target dynamics associated with $\langle \hat{\varphi}  \rangle_{\mu, n}$ for $\mu = 0, 1$ rotate at approximately the oscillator frequency in lab frame, for small fluxonum computational energy splittings. Therefore, the flux drive on the fluxonium should have the same frequency as the dressed oscillator, $\omega_d\sim\omega_\text{osc}$, to bring the desired evolution on resonance. 

\begin{figure}[t!]
    \centering
    \includegraphics[width = 0.45 \textwidth]{ 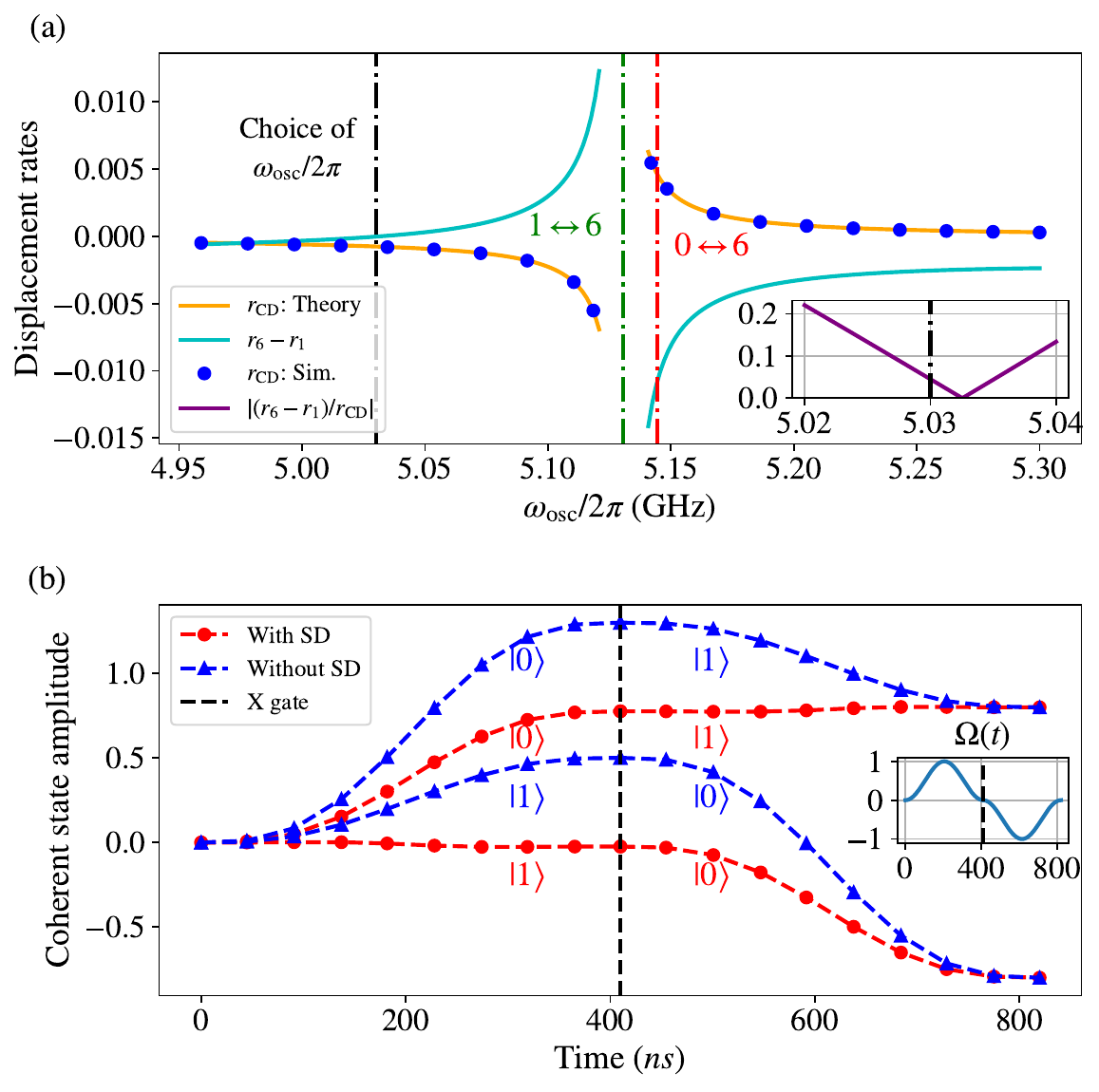}
    \caption{(a) Displacement rates with varying oscillator frequency for fluxonium $F_1$ in Tab.~\ref{tab:fluxonium_params}. The coupling strength is assumed to be $g/2\pi = 10$ MHz. The cyan and orange solid lines are obtained from perturbation theory predictions of the displacement rates in Eq.~\eqref{eq:disp_rate}. The blue dots are the displacement rates from numerical optimizations. The green, red and black dashed lines represent the frequency of fluxonium transitions $\ket{1}\leftrightarrow \ket{6}$, $\ket{0}\leftrightarrow \ket{6}$, and the choice of oscillator frequency, $\omega_{\text{osc}}/2\pi = 5.03$ GHz. The inset plot shows the ratio given in Eq.~\eqref{eq:rate_constraint} at the vicinity of choice of oscillator frequency. (b) The coherent state amplitude of the oscillator with the oscillator initialized in vacuum and the fluxonium in states $\ket{0}$ or $\ket{1}$. \rev{The amplitude is obtained from the coherent state with maximal overlap with the oscillator's state, $\max_{\alpha}\abs{\bra{\alpha}\hat{\rho}\ket{\alpha}}^2$, with $\rho$ being the oscillator state.} The state labels represent the fluxonium state. At the middle of the gate, a Pauli-X gate is applied to the fluxonium, flipping the states. The red circles (blue triangles) correspond to the oscillator dynamics with(out) selective darkening by having $c_{\text{SD}} = -r_1$ ($c_{\text{SD}} = 0$). The inset shows the pulse shape, $\Omega(t)$. \label{fig:demo_example}}
\end{figure}

To be more quantitative, we can define the displacement rate per unit drive strength conditioned on the fluxonium state. From first order perturbation theory, the unitless conditional displacement rates are 
\begin{eqnarray}
    r_k &:=& \frac{1}{\sqrt{n}}\bra{\phi_{k, n}} \hat{\varphi}\ket{\phi_{k, n-1}}\nonumber\\
    &\approx& ig\sum_{j\neq k} n_{k, j}\varphi_{j,k} \left(\frac{1}{\Delta_{k,j} - \omega_{\text{osc}}} - \frac{1}{\Delta_{k,j} + \omega_{\text{osc}}}\right), \label{eq:disp_rate}
\end{eqnarray}
where the fluxonium energy gaps $\Delta_{k, j} := E_j - E_k$, and $n_{k,j}$ ($\varphi_{k,j}$) are the matrix elements of the bare charge (flux) operator. For simplicity, we will refer to $r_k$ as the conditional displacement rate in the following discussions. The CD gate rate $r_{\text{CD}}$ follows naturally as
\begin{eqnarray}
    r_{\text{CD}} = r_1-r_0.
\end{eqnarray}
The values of $r_k$ can be numerically extracted from the coupled system Hamiltonian through optimization (See Appendix~\ref{sec:pert_theory_disp_rates} for details). In Fig.~\ref{fig:demo_example}(a), we focus on the specific example of fluxonium $F_1$ from Table.~\ref{tab:fluxonium_params}. The solid lines represent the theoretical predictions from Eq.~\eqref{eq:disp_rate}, and the blue dots represent the numerically extracted displacement rates, showing strong agreement with our perturbative treatment of the effective dynamics. Higher order effects in the perturbation theory could become relevant when the oscillator frequency is brought too close to of one of the fluxonium transition frequencies, e.g., $\Delta_{16}$, as the hybridization increases. As we elaborate in Sec.~\ref{sec:params_choice}, it is in general not desirable to operate in such a regime where the systems are strongly hybridized. 

Besides the desired dynamics associated with $r_{\text{CD}}$, additional spurious terms may become significant in our scheme under the fluxonium-oscillator hybridization. To improve the unitary gate fidelity, we employ techniques such as selective darkening and echoing, described in detail below.

\paragraph{Echoing.} Echoing is a common dynamical decoupling technique to cancel correlated errors. For example, within quantum computing, dynamical decoupling is applied in qubit-qubit CR gates~\cite{Jurcevic_2021} and qubit-oscillator gates~\cite{Eickbusch_2022_NatPhys}. When an oscillator is coupled to a qubit, the coherent error induced by spurious dispersive shifts can be detrimental if it is not properly compensated. At the middle of our gate, we apply a logical bit-flip on the fluxonium such that $\ket{0}$ flips to $\ket{1}$ and vice versa. Additionally, in the second half of the gate, the sign of the flux drive pulse is reverted. The pulse applied in Eq.~\eqref{eq:drive_Hamiltonian} has the form of
\begin{eqnarray}
        \Omega(t) = \sin^2\left(2\pi t/T\right)\label{eq:waveform}
\end{eqnarray}
for $0 \leq t\leq T/2$. For $T/2 \leq t^\prime\leq T$, $\Omega(t^\prime) = -\Omega(t^\prime-T/2)$. Here, $T$ is the gate time; see inset of Fig.~\ref{fig:demo_example}(b) for an illustration. To reduce leakage, the pulse should be adiabatic, and we numerically observe that choosing other similar waveforms, such as Gaussian envelopes, have negliglbe effect on the gate performance with sufficient ramp duration. For the following simulations, we choose the waveform in Eq.~\eqref{eq:waveform}. For simplicity, we assume an instantaneous, perfect X gate at $t = T/2$ to model the echoing, motivated by demonstrations of fast and high-fidelity fluxonium single qubit gates~\cite{Zhang_2021, PhysRevLett.129.010502, PRXQuantum.3.037001}.

\paragraph{Selective darkening.} Selective darkening (SD)~\cite{de_Groot_2010, de_Groot_2012, PhysRevApplied.18.034063} is a technique originally developed for multi-qubit gates. In the context of controlling an oscillator, the core motivation is that the existence of dispersive shift leads to variations in the oscillator frequency conditioned on the fluxonium state. Therefore, it is not possible to simultaneously and resonantly drive the oscillator when the fluxonium is in a superposition of different states. To achieve SD, we add an unconditional, linear drive at the frequency of the oscillator corresponding to the fluxonium state $\ket{\mu}$, to generate the unitary
\begin{eqnarray}
    \hat{U}^{\prime}_{\text{CD}, \alpha} = \vert \mu\rangle \langle \mu \vert \otimes  \hat{I}+ \vert \nu\rangle \langle \nu \vert\otimes \hat{D}_{\alpha},
\end{eqnarray}
where $\mu\neq \nu$ and $\mu, \nu$ are computational states. These two gates are are equivalent up to an unconditional displacement, but the new unitary effectively darkens the oscillator when the fluxonium is in $\ket{\mu}$. The drive frequency can now be unambiguously chosen to be the oscillator frequency conditioned on fluxonium in $\ket{\nu}$. In particular, to achieve $\hat{U}^{\prime}_{\text{CD}, \alpha}$, we require that the amplitude of the SD drive is
\begin{eqnarray}
    c_{\text{SD}} = -r_{\mu}.
\end{eqnarray}
\rev{This condition ensures that the oscillator dynamics are selectively darkened at all times during the gate. To drive the oscillator on-resonance, the drive frequency $\omega_d$ is naturally the oscillator frequency conditioned on the the fluxonium in $\ket{\nu}$.}

An addition of an $\text{X}$ gate for ECD gates ensures that the resonator will not be displaced for half of the gate duration. Indeed, an additional benefit of SD for the ECD gate is that it reduces the average photon occupation number during the gate compared to the gate without SD. This is in turn desirable to suppress both coherent and incoherent errors, such as spurious Kerr Hamiltonians and oscillator dephasing noise. 

Equipped with the techniques above, we demonstrate the pulse shape and oscillator dynamics in Fig.~\ref{fig:demo_example}(b). The oscillator is initialized in vacuum, and we run a time-domain simulation of the full Hamiltonian $\hat{H}_{sys}$ including the drive $\hat{H}_d$. For various times $t$, we fit the oscillator state conditioned on fluxonium states to a coherent state to extract the oscillator state amplitudes. Without SD, the blue line shows conditional displacements to amplitudes of $0.8(-0.8)$ over the full gate duration of 800ns, for the fluxonium $\ket{1}$($\ket{0}$) state. The X gate applied at the center at $\sim 400$ns inverts the fluxonium states and, accordingly, the direction of the oscillator displacements. In the red solid line, we have chosen to perform SD and darken the oscillator dynamics when the fluxonium is in $\ket{1}$. As a result, it is clear that during the gate, the oscillator is only displaced when the fluxonium is in $\ket{0}$. As a comparison, the blue lines represent the oscillator dynamics without SD, and the oscillator is displaced when for both fluxonium states $\ket{0}$ and $\ket{1}$. 

Moreover, since we have applied a bit-flip at the middle of the gate, the oscillator states always stay in a low photon space during the gate compared to the case without SD. While we have not yet optimized the circuit or control tone, it is clear that fast, conditional displacements are possible with typical parameter values from the literature.

\begin{table*}
    \centering
    \begin{tabular}{p{2cm}<{\centering} p{2.5cm}<{\centering} p{2.5cm}<{\centering} p{2.5cm}<{\centering} p{2.5cm}<{\centering} p{2.5cm}<{\centering}}
        \hline \hline
        Fluxonium label & $E_{J}/h$ (GHz) & $E_{L}/h$ (GHz) & $E_{C}/h$ (GHz) & $\Delta \Phi/\Phi_0$ & $\omega_f/2\pi$ (MHz)\\
         \hline 
         $F_1$ & 3.395 & 0.132 & 0.479 & 0.0 & 14\\
         $F_2$ & 3.27 & 0.125 & 0.462 & 0.0 & 13\\
         $F_3$ & 5.71 & 0.59 & 1.3 & 0.08 & 1600\\
         $F_4$ & 4.0 & 1.0 & 1.0 & 0.017 & 734\\
         \hline \hline 
    \end{tabular}
    \caption{Summary of the fluxonium parameters. The parameters $F_1$, $F_3$, $F_4$ are adopted from Ref.~\cite{Zhang_2021}, Ref.~\cite{PhysRevX.13.031035}, and Ref.~\cite{PRXQuantum.3.037001} respectively. Parameters $F_2$ is obtained through varying the parameters of fluxonium $F_1$. The fluxonium frequency is the energy gap between the computational states, $\omega_f = \Delta_{0,1}$.}
    \label{tab:fluxonium_params}
\end{table*}

\section{Effects and choices of system parameters\label{sec:params_choice}}

In this section, we detail the potential trade-offs in the choice of system parameters, including the energies of the fluxonium and the oscillator, the coupling strength, and the drive strength. We also outline a series of criteria, which is based only on the fluxonium spectrum and its charge and flux matrix elements, that narrows down the promising range of values for all gate parameters and oscillator frequency. Our method for searching through the parameter space is computationally efficient since it avoids the full time-domain simulation across the high-dimensional parameter space and full diagonalization of the coupled system. Here, we stress that while our goal is to provide a set of guiding principles and intuitions towards parameter selections, they are by no means optimized. Moreover, the choice of metric depends strongly on the application of interest. 

To make our discussions more concrete, in each subsection we start with general discussions and then focus on a set of fluxonium parameters, $F_1$ in Table.~\ref{tab:fluxonium_params}, as an example. \rev{Nevertheless, because our analysis methodology is specified in terms of the fluxonium spectrum and matrix elements, it is fully general and not restricted to any specific fluxonium parameters.} As we walk through this section, we settle down a set of working gate parameters one-by-one, which are adopted in the analysis of gate errors in Sec.~\ref{sec:error_budget}.

\subsection{Fluxonium parameters}

The fluxonium parameters are $\set{E_{J}, E_L, E_C}$ and the external flux, $\Delta \Phi$. In this section, we provide intuition on each parameter's significance and operational choice. Recall from Eq.~\eqref{eq:disp_rate} that the gate is activated through the hybridization of computational and non-computational states, labelled by $\ket{\mu}$ and $\ket{i}$ respectively. Note that while the two states could in principle both be computational states, it is not desirable since for fluxonium, the frequency is generally low, i.e. on the level of hundreds of MHz or even lower, and it could be hard to bring the oscillator to such frequencies. Additionally, the difference in the displacement rates $r_{0}$ and $r_{1}$ is what gives rise to the gate speed. Therefore, the errors posed by mismatched displacement rates, which we elaborate in Section.~\ref{sec:error_budget}, would be outstanding.

There are three key factors when we consider the fluxonium qubit. Firstly, the flux and charge matrix elements, $\bra{i}\hat{\varphi}\ket{\mu}$ and $\bra{i}\hat{n}\ket{\mu}$ set the effective gate speed and should not be small. In simulations, the matrix elements are calculated numerically. In later subsections we outline conditions on the gate and system parameters, and having unfavorable matrix elements could lead to failure in satisfying those conditions. Secondly, if one considers open system simulations and fault-tolerance, \rev{it is desirable to have a suppressed auxiliary qubit bit-flip probability over the gate duration. One natural choice to achiveve this would be to operate a heavy fluxonium off sweet spot. However, fluxoniums operated on sweet spot could benefit from shorter gate times or higher unitary gate fidelities while having a decent bit-flip rate. Therefore, we consider fluxoniums both on and off sweet spots, leaving the search for the optimal operating parameters to future work.} The last consideration is whether the fluxonium parameters are easy to fabricate in practice. Therefore, the fluxonium parameters chosen in Tab.~\ref{tab:fluxonium_params} mostly come from existing literatures.

For example, for fluxonium $F_1$ in Table.~\ref{tab:fluxonium_params}, the qubit is a heavy fluxonium, with $E_J\gg E_C$, operating on sweet spot with a frequency of $\omega_{f}/2\pi = 14$MHz. In Fig.~\ref{fig:demo_example}(a), we sweep the oscillator frequency, $\omega_{\text{osc}}/2\pi$, over the range of $[4.95, 5.30]$ (GHz) and plot the displacement rates. In this frequency range, the two relevant fluxonium transitions are $\ket{0}\leftrightarrow \ket{6}$ and $\ket{1}\leftrightarrow \ket{6}$, which are represented by green and red dashed lines, respectively. Since the qubit is on the sweet spot, the transition $\ket{0}\leftrightarrow \ket{6}$ is parity forbidden with vanishing matrix elements $\hat{\varphi}_{0,6} = \hat{n}_{0,6} = 0$. This can be seen in Fig.~\ref{fig:demo_example}(a) where the displacement rates are not affected when the oscillator frequency is close to the red dashed line. On the contrary, the magnitude of the CD displacement rate $r_{\text{CD}}$ increases drastically close to the transition frequency of $\ket{1}\leftrightarrow \ket{6}$, through increased hybridization, setting the transition through which we can activate the gate (in this case, $\mu = 0$ and $i = 6$).

\subsection{Oscillator frequency\label{sec:osc_freq_criteria}}

For the gate to be activated through a transition $(\mu, i)$, the oscillator frequency must stay in the vicinity of the transition frequency, $\Delta_{\mu, i}$, for a significant hybridization and gate rate. By considering two constraints, we can bound the interval of oscillator frequency that gives valid working parameters. 

The oscillator frequency, $\omega_{\text{osc}}$, should refrain from being too close to $\Delta_{\mu, i}$ to avoid mismatched displacement rates between $r_i$ and $r_\mu$. In the limit where the oscillator frequency approaches the transition frequency, we find $r_i = -r_\mu$, which can be derived straightforwardly from Eq.~\eqref{eq:disp_rate}. As we will elaborate in Sec.~\ref{sec:error_budget}, such a mismatch (due to opposite signs between $r_{i}$ and $r_{\mu}$) could lead to errors and a compromised gate rate. Therefore, the displacement rates should satisfy 
\begin{eqnarray}
    \abs{r_i - r_\mu}/\abs{r_{\text{CD}}}\leq C_r \label{eq:rate_constraint}
\end{eqnarray}
with $C_r$ being a constant, which is given a heuristic value  of $C_r = 0.2$. Such a condition effectively puts a constraint on $\omega_{\text{osc}}$. For example, in Fig.~\ref{fig:demo_example}(a), when the oscillator frequency approaches the transition $\ket{1}\leftrightarrow \ket{6}$, the displacement rate difference, $r_6 - r_1$, increases rapidly.

Meanwhile, it is not desirable to have $\omega_{\text{osc}}$ being far from $\Delta_{\mu, i}$, in which case the gate would require strong coupling and strong drives. The exact limit on the strength of the coupling and drives depend on the experimental conditions, circuit requirements, and the application at hand. For clarity, we follow the constraint that one and only one of the transition should give dominant contribution to the displacement rate, where the contribution to the displacements rates from each transition is obtained through each term in the summation in Eq.~\eqref{eq:disp_rate}. However, there is no fundamental limitation towards a cross-resonance gate stimulated by multiple transitions, and this is a constraint for the simplicity of analysis. 

For any valid transition $(\mu, i)$, the two constraints described above provides a bounded range of $\omega_{\text{osc}}$. The process of determining the oscillator frequency can be done through an efficient numerical scan given the fluxonium spectrum and matrix elements. With the bounds described, we select the oscillator frequency for fluxonium $F_1$ to be $\omega_{\text{osc}}/2\pi = 5.03$ GHz, which corresponds to a conservative ratio of $\abs{r_i - r_\mu}/\abs{r_{\text{CD}}} \approx 0.05$, and $\Delta_{16} - \omega_{\text{osc}} = 100.4$MHz. 

\subsection{Coupling strength}

With finite hybridization between the fluxonium and oscillator, additional spurious effects such as inherited self-Kerr nonlinearities may be introduced. Here, we describe the associated limitations on hybrization as a requirement in coupling strength, $g$:
\begin{eqnarray}
    g\sqrt{N}\abs{\frac{\bra{i}\hat{n}\ket{\mu}}{\Delta_{i,\mu}-\omega_{\text{osc}}}}\leq C_{g},\label{eq:coupling_bound}
\end{eqnarray}
where $C_g$ is a constant set to $0.1$ as a heuristic. Here, $N$ represents the index of the highest-energy fock level of interest. Intuitively, the left-hand side of Eq.~\eqref{eq:coupling_bound} roughly represents the maximum amplitude of the hybridization between $\ket{\mu}\ket{n}$ and $\ket{i}\ket{n-1}$ with capacitive coupling. With $N = 10$, fluxonium $F_1$ parameters, and the fixed oscillator frequency, the constraint above gives $g/2\pi\leq g_{\text{max}}/2\pi = 47.8$ MHz. In simulations, we choose rather conservative coupling strengths. For instance, in Sec.~\ref{sec:error_budget}, the coupling strength of choice is $g/2\pi = \frac{g_{\text{max}}/2\pi}{3} = 15.9$MHz. As a result, we have a dispersive shift $\chi/2\pi = 26.1$kHz and a self-Kerr of $K/2\pi = 0.783$kHz. Since the gate design includes echoing, the dispersive shift does not introduce error. However, the self-Kerr and strong nonlinear hybridization can degrade the performance of an ECD gate, hence the heuristic constraint on $g$ in Eq.~\eqref{eq:coupling_bound}. The gate times we have obtained are generally under 1$\mu$s, so such extent of hybridization is tolerable.

\subsection{Drive strength and gate time}

The final tunable parameter is the drive strength, $A$, which then determines the gate duration. In practice, there could be requirements on the gate duration when considering factors like incoherent error rates. \rev{For example, a strong flux drive for a fast gate could induce stray fields that lead to degraded gate fidelity. Such external requirements will depend on the particulars of the experimental platform, so here instead we constrain the drive strength by restricting leakage errors. As a result, the analysis below provides estimates on the shortest gate time without being affected by leakage errors.}

The drive strength is effectively upper-bounded to reduce population leakage to noncomputationl states. In the limit of weak drive and the oscillator frequency being in the vicinity of a single fluxonium transition, we expect that the gate is activated predominantly by one transition $(\mu, i)$, and the non-computational state most likely to be occupied is $\ket{i}$. Thus, we bound the drive strength as
\begin{eqnarray}
    A\abs{\frac{\bra{i}\hat{\varphi}\ket{\mu}}{\Delta_{i,\mu}-\omega_{\text{osc}}}}\leq C_{A},
\end{eqnarray}
where we commonly set $C_A = 1$ for an approximate order-of-magnitude estimate. For example, with the given fluxonium parameter $F_1$ of Table.~\ref{tab:fluxonium_params} and the oscillator frequency determined through Sec.~\ref{sec:osc_freq_criteria}, an upper bound for the amplitude is $A/2\pi\leq A_{\text{max}}/2\pi= 1.6$ GHz. Equivalently, a lower bound on the gate time, $T_{\text{min}}$, can then be estimated as 
\begin{eqnarray}
    L \lesssim \frac{1}{2}A_{\text{max}}r^{\text{max}}_{\text{CD}}\int_{0}^{T_{\text{min}}} \Omega(t)dt,
\end{eqnarray}
where $L$ is the target displacement length, and $r^{\text{max}}_{\text{CD}}$ is the CD gate rate when the coupling strength takes the maximum allowed coupling strength, $g = g_{\text{max}}$, in Eq.~\eqref{eq:disp_rate}. Our description is simply an order-of-magnitude estimation, since in practice some effects, such as mismatched displacment rates, could compromise the gate rate $r_{\text{CD}}$. With fluxonium $F_1$ and system parameters determined through previous sections, the estimated gate duration is $T_{\text{min}} \sim 300ns$. Fluxonium $F_1$ has a decay lifetime of $T_1 = 315\mu $s~\cite{Zhang_2021}, which lead to $\sim 10^{-3}$ probability of auxiliary qubit decay per CD gate. It is worth noticing that fluxonium $F_1$ is on sweet spot, and it has been shown that with off sweet spot fluxonium, $T_1$ times can be extended to $10$ms~\cite{PhysRevLett.120.150504}.

\section{Error budget\label{sec:error_budget}}  

In this section, we provide an analysis of the coherent gate error. Since we are working with oscillators, which have infinite-dimensional Hilbert spaces, it is critical to set a Fock space cutoff dimension, $N_{\text{max}}$, to make simulations manageable. Equally importantly, we define an \textit{input} subspace-of-interest with projector $\hat{P}$, which restricts the unitary fidelity to a further truncated space that the CD gate is applied to. Physically, the input subspace-of-interest represents the space that encodes information \textit{before} the CD gate. Such a notion provides a sensible measure of the gate's performance because most practical applications do not utilize the whole space, and the coherent and incoherent errors can be amplified at the subspace of high excitations. Since only the computational space of the fluxonium stores information, we have $\hat{P} = \sum_{i=0}^1\sum_{n = 0}^{N_{\text{in}}}\dm{\phi_{i, n}}$. The oscillator state during the CD gate should be well contained within the Fock state truncate, implying $N_{\text{in}}\leq N_{\text{max}}$. Unless specified otherwise, we set $N_{\text{in}} = 10$. $N_{\text{max}}$ depends on parameters such as the target displacement length and $N_{\text{in}}$. In our simulations, we set $N_{\text{max}} = 45$ and observe that the results are not significantly affected with an increased $N_{\text{max}}$.

The target gate we choose is an eigenbasis ECD gate, as given in Eq.~\eqref{eq:original_CD}, with displacement length $\alpha = 1.6$, which is a decent gate size while still being numerically tractable. The displacement length is defined such that a displacement operator of length $\alpha$ displaces an oscillator in vacuum state into the coherent state $\ket{\alpha}.$ As a reference, to implement the small-Big-small protocol~\cite{Royer_2020}, the longest required CD gate is of displacement length $\sqrt{2\pi}\approx 2.5$. With the fixed displacement length and system parameters, we scan the gate performance over a range of gate times by varying the drive strength. Note that there are also certain \textit{free} operations that we can apply to compensate the gate, such as oscillator and fluxonium phase rotations, which can be tracked in software. Since the gate scheme includes echoing, unconditional displacement is in principle not required. See Appendix~\ref{sec:gate_details} for the relevant simulation details. Suppose the target gate is $\hat{U}_t$, the actual dynamics is $\hat{U}$, and the projector onto the subspace-of-interest is $\hat{P}$, the gate fidelity has the form of~\cite{Pedersen_2007}
\begin{eqnarray}
        F = \frac{1}{N_p(N_p+1)}\left[\Tr{M^\dagger M} + \abs{\Tr{M}}^2\right]\label{eq:gate_infidelity}
\end{eqnarray}
where $M:= \hat{P}\hat{U}_t \hat{U} \hat{P}$, and $N_p = \Tr \hat{P} = 2N_{\text{in}}$. 

\begin{figure}[t!]
    \centering
    \includegraphics[width = 0.45 \textwidth]{ 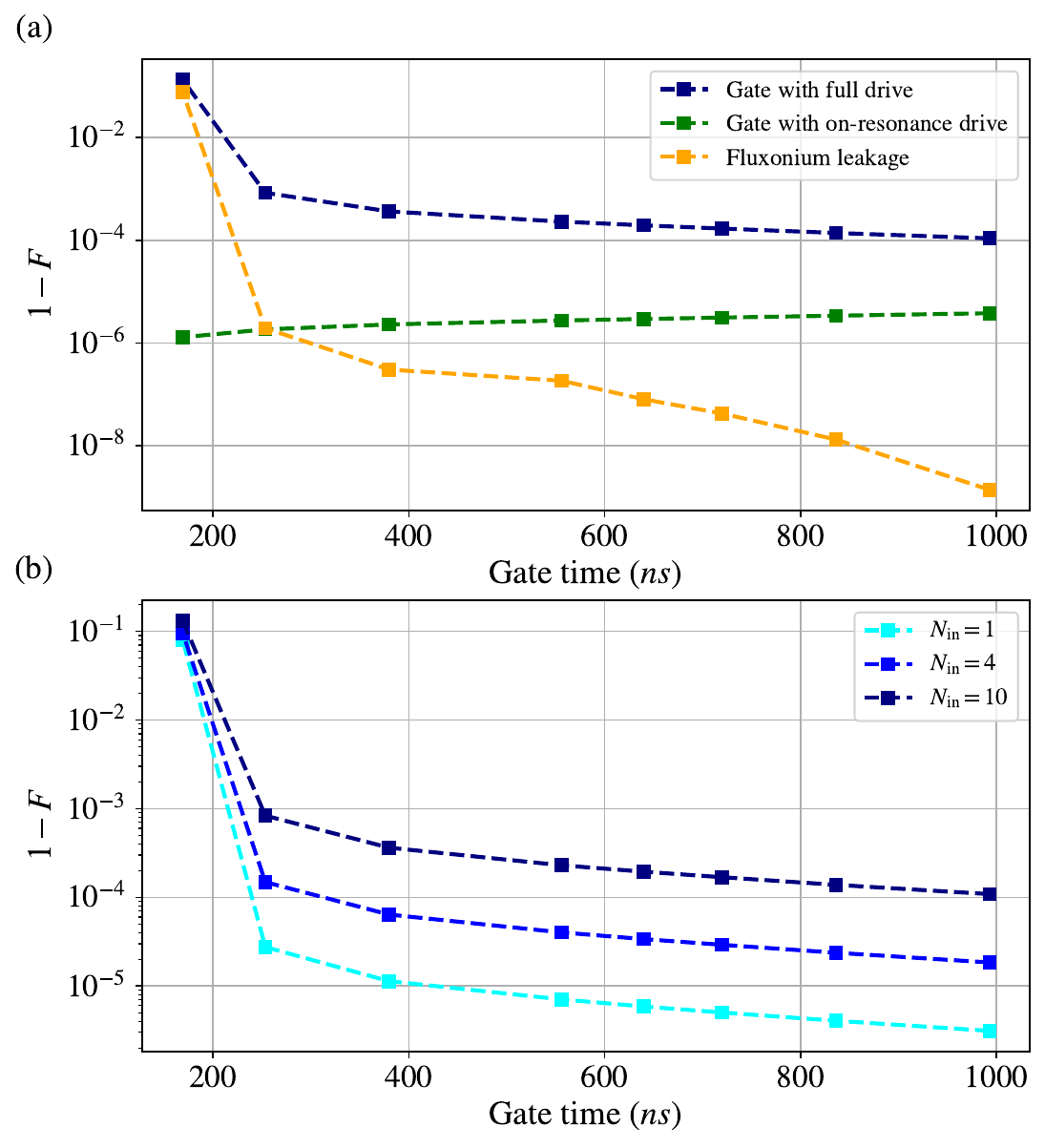}
     \caption{(a) The gate infidelity for fluxonium $F_1$ in Tab.~\ref{tab:fluxonium_params}, oscillator frequency $\omega_{\text{osc}}/2\pi = 4.86$ GHz, coupling strength $g/2\pi = 8$ MHz. Blue (orange) squares represent gate driven by the full (on-resonant) drive Hamiltonian. Green squares represent the leakage when the fluxonium is decoupled from the oscillator. The red circles represent the average state infidelity for the task of cat state preparation. The target gate achieved by the blue squares is an ECD gate, as given in Eq.~\eqref{eq:original_CD}, with displacement length $\alpha = 1.6$. (b) Gate infidelity for varying sizes of input subspace-of-interest. $N_{\text{in}}$ represents the number of Fock levels contained in the input subspace-of-interest. The case with $N_{\text{in}}=1$ corresponds to the task of cat state preparation.}
    \label{fig:error_budget}
\end{figure}

We emphasize that the metric depends greatly on the application in mind, which could lead to, for example, significant difference in the subspace-of-interest. To give a sense of the potential difference in various metrics, in Fig.~\ref{fig:error_budget}(b) we present the gate fidelity for various input subspace-of-interest cutoffs. Clearly, with the oscillator restricted to lower energy states, the coherent errors improves significantly. In particular, the setting with the oscillator initialized in vacuum, $N_{\text{in}}=1$, is applicable for many tasks. One such example is cat state preparation: the CD gates are applied to $\ket{\pm}\ket{0}$, and conditioned on the fluxonium being measured in $\ket{\pm}$ ($\ket{\mp}$), the oscillator is prepared in the even (odd) cat states. Another example is for reading out qubit states with the oscillator.

The drive Hamiltonian $H_{\text{d}}$, as given in Eq.~\eqref{eq:drive_Hamiltonian}, is driven at the oscillator frequency. Therefore, in the interaction picture, terms in the flux drive Hamiltonin, $\hat{\varphi}$, that are proportional to $\ket{\phi_{\mu, n}}\bra{\phi_{\mu, n+1}} + \text{h.c.}$ are resonant for any $\mu$ and $n$. We define these terms as the on-resonance drive. When these terms are excluded, the residual drive Hamiltonian is referred to as the off-resonance drive. In ideal scenarios, the on-resonance drive should only contain terms that generate the conditional displacement, and the off-resonance drive should vanish due to rotating wave appriximation. It is exactly the deviations from these two approximations that lead to gate errors. In Fig.~\ref{fig:error_budget}(a), we adopt a divide-and-conquer strategy, where we divide the drive operator elements into on-resonance and off-resonance terms to isolate and identify the error sources.

\subsection{On-resonance terms}

The on-resonance terms refer to the drive Hamiltonian matrix elements proportional to $\ket{\phi_{\mu, n}}\bra{\phi_{\mu, n-1}} + \text{h.c.}$ for any $\mu, n$. To first order, as shown in Eq.~\eqref{eq:disp_rate}, the coefficients of the corresponding flux operator elements are proportional to the bosonic enhancement factor, $g\sqrt{n}$, as desired by the coefficients of $\hat{U}_0 \sbkt{\hat{a} + \hat{a}^\dagger}\hat{U}_0^\dagger$. However, higher-order effects could lead to deviations from such a scaling. To leading order, they could be approximately proportional to $\sbkt{g \sqrt{n}}^3$, which can be verified by the numerical fits given in Appendix~\ref{sec:pert_theory_disp_rates}. While these deviations are suppressed for weakly coupled systems, they scale unfavorably with increasing oscillator energies and contribute to the behavior shown in Fig.~\ref{fig:error_budget}(b).

We isolate the effect of the on-resonance errors through simulating the ECD gate with a modified drive Hamiltonian, in which only on-resonance terms remain. In Fig~\ref{fig:error_budget}(a), the green square represents the gate infidelity with the on-resonance drive. The resulting infidelity should only have a strong dependence on the coupling strength but not on the drive strength. Such a feature is clear from Fig~\ref{fig:error_budget}(a), where the infidelity curve remains approximately constant over a range of gate times. The infidelity is considerably lower than the infidelity of the gate driven by the full drive Hamiltonian at short gate times. Only at long gate times where the off-resonance error is suppressed by the drive strength reduction could the on-resonance terms be a limiting factor.

\subsection{Off-resonance terms}

The off-resonance terms are mostly contributed by the flux drive in $H_{\text{d}}$, and these terms cause some non-computational fluxonium levels to be populated during the gate. For example, if we assume the gate is activated by the pair of computational and non-computational states $(\mu, i)$, the oscillator frequency is brought decently close to the transition frequency $\Delta_{\mu, i}$. Therefore, while the transition $\ket{\mu} \leftrightarrow \ket{i}$ is off-resonant, $\omega_{\text{osc}}-\Delta_{\mu, i}$ could be comparable to the drive strength, thus breaking the rotating wave approximations. In Fig~\ref{fig:error_budget}(a), the blue square represents the gate infidelity when its driven by the full drive Hamiltonian, and the performance gap between the gate driven by the full drive and the on-resonant drive represents the errors brought by the off-resonance terms (including any interplay between on-resonant and off-resonant terms). Physically, the consequences of the off-resonance terms include leakage and mismatched displacement rate, which we detail below.

\subsubsection{Leakage}

Recall from Sec.~\ref{sec:model_gate} we motivated the adiabatic pulse form in Eq.~\ref{eq:waveform} by leakage reduction. However, with increasing drive strength, the adiabaticity is no longer sufficient. Since the logical information is encoded in the full space of the oscillator, the leakage is defined as any remaining population in the non-computational levels of the fluxonium at the end of the gate. While there are other potentially useful techniques like derivative removal by adiabatic gate (DRAG)~\cite{li2023suppression, PhysRevA.88.062318, Theis_2018}, we do not explore them here and leave it for future work.

To isolate the leakage error, a straightforward method is to compute the leakage population after acting the unitary $\hat{U}$ on the subspace-of-interest $\hat{P}$. Since the leakage is only defined with respect to the fluxonium computational levels, the leakage can be simulated with the fluxonium alone. In Fig~\ref{fig:error_budget}(a), the orange squares represent the leakage of an isolated fluxonium driven by a flux with the same drive strength and pulse shape. It is indeed clear that at strong drive and short gate times, the leakage can be explained by such a simplified model. This also opens the door to the possibility of bounding the drive strength from above with relatively efficient simulations.

\subsubsection{Mismatched displacement rate}

Leakage error explains the error budget at strong drives. At weaker drive strengths, we were not able to find a simplified model to explain the residual off-resonance error, i.e.~the infidelity gap between the full drive and on-resonance drive gates. However, one mechanism that does contribute is the mismatched displacement rate.

Suppose we follow the assumption that one transition, $\sbkt{\mu, i}$, contributes dominantly to the gate rate, the mismatched displacement rate refers to the rate difference $\abs{r_i - r_\mu}$. The rise of such an effect can be understood by modeling the fluxonium as a two-level system composed of $\ket{\mu}$ and $\ket{i}$. As we adiabatically bring up the flux drive, the fluxonium state will remain in the time-dependent eigenstate $\ket{\tilde{\mu}}$. Here, $\ket{\tilde{\mu}}$ is composed of both $\ket{\mu}$ and $\ket{i}$, which induces displacement in the oscillator at different rates. Such a realization brings modifications to the displacement rates given in Eq.~\eqref{eq:disp_rate}, which are derived by focusing on the static matrix elements. The effective time-dependent displacement rates of the fluxonium initialized in $\ket{\mu}$ can be derived to be
\begin{eqnarray}
        r_{\mu}(t) = r_\mu + (r_i - r_\mu)\abs{\frac{A\Omega(t)}{\Delta_{\mu, i} - \omega_{\text{osc}}}\bra{i}\hat{\varphi}\ket{\mu}}^2.\label{eq:mod_disp_rate}
\end{eqnarray}
Therefore, one effect of the mismatched displacement rate is that it could lead to a modified displacement rate. In the limit where the oscillator frequency is sufficiently close to $\Delta_{\mu, i}$, $r_\mu \approx -r_i$, and it is likely that such a modification will compromise the gate rate. Moreover, since the oscillator dynamics conditioned on the fluxonium state is different, the oscillator dynamics generally would not be a perfect displacement when the fluxonium state is adiabatically brought back to the initial computational state at the end of the gate (see Appendix.~\ref{sec:mismatch_disp_rates} for more details). Therefore, the above effects motivate\kn{\sout{s}} the constraint set by Eq.~\eqref{eq:rate_constraint} to suppress the effects of the mismatched displacement rate.

\begin{figure*}
    \centering
    \includegraphics[width = 1.0 \textwidth]{ 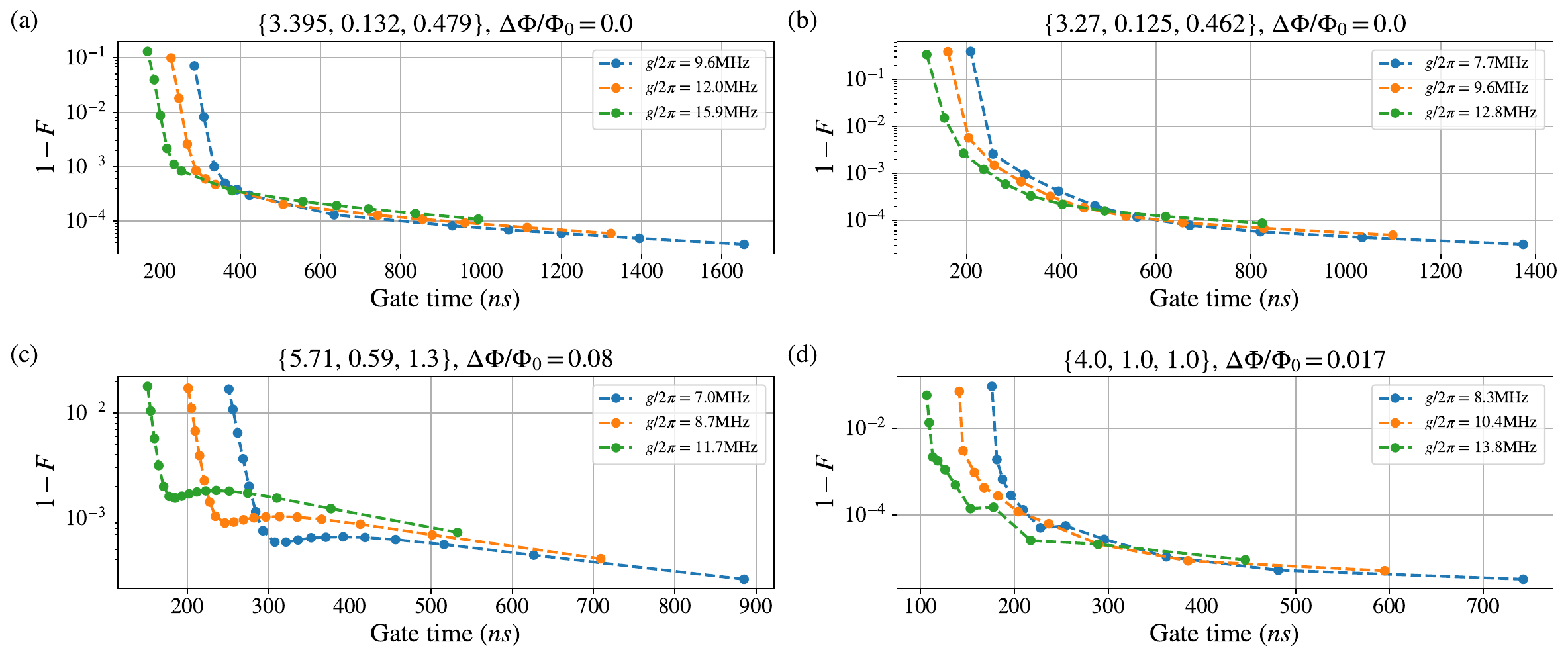}
     \caption{A summary of the gate performance under various coupling strength for fluxonium parameters (a) $F_1$, (b) $F_2$, (c) $F_3$, and (d) $F_4$. The fluxonium energy parameter sets are labelled in the order of $\left\{E_J/h, E_L/h, E_C/h\right\}$, all in units of GHz. The oscillator frequencies chosen for (a-d) are $\omega_{\text{osc}}/2\pi = 5.03, 4.86, 6.89, 6.0$ (GHz) respectively. The target gate is an ECD gate with displacement length $\alpha = 1.6$. }
     \label{fig:survey}
\end{figure*}

\section{Survey of gate performance for various fluxonium parameters\label{sec:survey}}

As elaborated in Sec.~\ref{sec:params_choice}, we have developed a relatively systematic procedure for selecting the resonator frequency based on the fluxonium energy parameters. In this section, we demonstrate the robustness and wide applicability of our gate design and parameter selection procedure. The target CD gate displacement length is set to be $\alpha = 1.6$ with $N_{\text{in}}=10$ as before. For each parameter set, we examine a few coupling strength choices that satisfy the constraints set in Eq.~\eqref{eq:coupling_bound}. Then, we vary the drive strength to achieve the target displacement length for a range of gate times. The set of fluxonium energy parameters we survey are given by Tab.~\ref{tab:fluxonium_params}, where the parameters are mostly adopted from existing literatures to ensure they are realizable with current fabrication techniques. Among the four fluxonium parameter sets, we have chosen half to be on sweet spot and the rest to be off sweet spot to demonstrate that our gate scheme is viable for both scenarios. In Fig.~\ref{fig:survey}, we present a summary of the gate performance. \rev{The drive strength required can be characterized by the external flux modulation. To give an estimate on the magnitude required, for $F_{1,2,3,4}$ to achieve a gate time of around $0.8\mu s$, the required maximum flux modulations are $0.57\Phi_0, 0.62\Phi_0, 0.068\Phi_0, 0.033\Phi_0$. To put the number into context, tuning the external flux over half flux quantum, $0.5\Phi_0$, covers the minimum and maximum fluxonium frequency achievable by controlling the external flux. More details and data on the drive strength are given in the appendix.}

\rev{We reiterate that the use of fluxonium for the auxiliary qubit is motivated by the potential for lower bit-flip rates in comparison to other commonly use qubits such as transmons. Since fluxonium can exhibit long $T_1$ both on and off sweet spot, we consider both cases here. In the appendix, we include the simulations of a heavy fluxonium, $F_1$ in Tab.~\ref{tab:fluxonium_params}, operated off sweet spot and achieving unitary error as low as $5\times 10^{-4}$ with $2\mu s$ gate time. At the external flux chosen, the fluxonium's $T_1$ has been experimentally measured to be around $4\mu s$~\cite{Zhang_2021}. 
}

Sec.~\ref{sec:error_budget} discussed various error sources of the gate. One of the features is that when leakage error dominates at strong drive strength, it can be estimated through simulating a decoupled fluxonium qubit alone. Such a feature is apparent from Fig.~\ref{fig:survey}: since the coupling strength affects the displacement rates, the effect of an increase in the coupling strength is roughly to translationally shift the infidelity curves towards short gate time. At the limit of weak drive, the dominating error source switches from off-resonance terms to on-resonance terms. Thus, at long gate times, the higher the coupling strength, the larger the infidelity.

Overall, it is clear that for most of the fluxonium parameters, the protocol devised in Sec.~\ref{sec:params_choice} offers valid choices of oscillator frequency, coupling strength, and drive strength such that the gate can be performed with infidelity less than $10^{-3}$ within $300 ns$. Moreover, one of the optimal performances, a gate infidelity of $5\times 10^{-5}$ with a gate time of $250 ns$, is achieved by fluxonium $F_4$. However, we stress again that our gate parameters are not necessarily fully optimized, and the main goal of our parameter selection procedure is to narrow down the parameter space and provide intuition on the role of each parameter.

\begin{figure}[t!]
    \centering
    \includegraphics[width = 0.45 \textwidth]{ 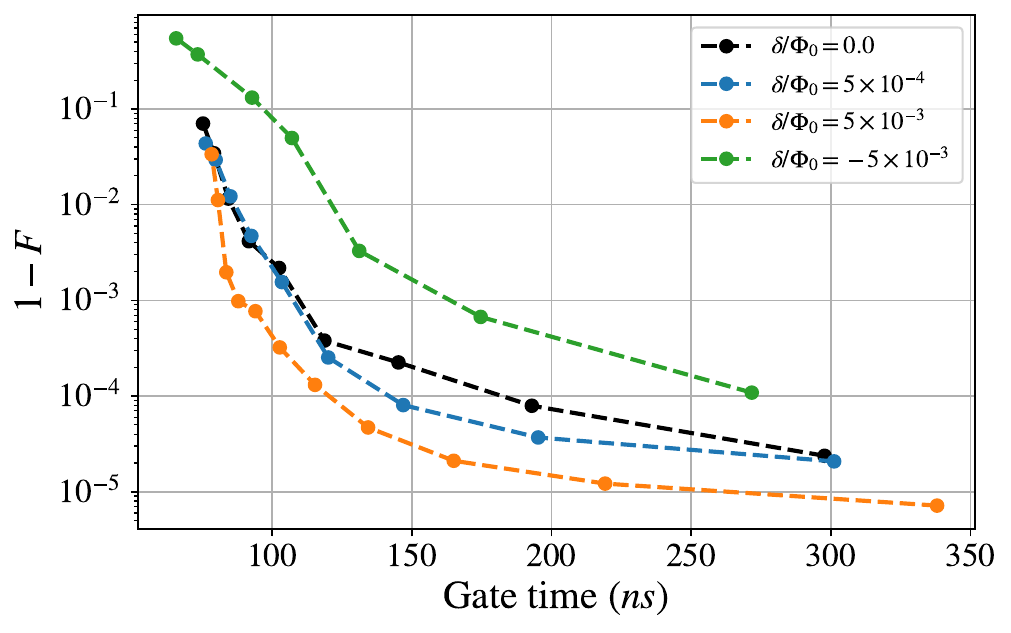}
     \caption{Robustness against external flux deviations. For all data points, the fluxonium energy parameters are taken to be the same as the parameter set $F_4$ in Tab.~\ref{tab:fluxonium_params} with $\omega_{\text{osc}}/2\pi = 6.0$ GHz, and the coupling strength is fixed to $g/2\pi = 20.8$ MHz. The external flux are varied and benchmarked against the external flux of parameter set $F_4$, where $\Delta \Phi_{F_4} = 0.017 \Phi_0$. The external flux deviation is defined as $\delta = \Delta \Phi - \Delta\Phi_{F_4}$.}
    \label{fig:flux_fluctuation}
\end{figure}

In realistic settings, the system parameters might deviate from the designed values. One reason for this could be the uncertainty in fluxonium fabrication processes that makes precise parameter targeting challenging. This motivates our choice of fluxonium parameters $F_2$ in Tab.~\ref{tab:fluxonium_params}. The parameters $F_2$ represent a random relative deviation of approximately $5\%$ compared to parameters $F_1$, which are parameters of real fabricated devices~\cite{Zhang_2021}. It is clear from Fig.~\ref{fig:survey}(a) and (b) that even if the fluxonium parameters are varied, we can still determine a valid oscillator frequency to achieve gates with decent performance. \rev{Worth noting, our scheme of controlling each oscillator with a fluxonium is fully sufficient for architectures based on multiple oscillators. The interactions between oscillators can be realized through other active couplers such as SQUID (superconducting quantum interference device). Such architectures have been proposed in recent works~\cite{lemonde2024hardwareefficientfaulttolerantquantum} and can directly adopt the gate scheme we propose in the current work. Going a step further, it could be desirable to be able to control multiple oscillators with a single fluxonium. However, since the gate rate and performance depends on the oscillator frequency, it could be challenging to find a scalable frequency allocation for such architectures, and we leave explorations along this direction to future work.}

Another potential concern is the fluctuation in the external flux. In Fig.~\ref{fig:flux_fluctuation}, fluxonium $F_4$ is taken as an example, where all other system energy parameters are fixed to be the same but the external flux is varied. Such a setup corresponds to the case when the external flux fluctuation is low-frequency, which is dominant for $1/f$ noise~\cite{10.1063/1.98041, Koch_1983_JLTP}. Since the displacement rates have changed, the gate parameters, such as the gate time, would need to be recalibrated, but it is clear that the gate's performance remains satisfactory. For the case where the external flux is increased by $5\times 10^{-3}\Phi_0$, the gate infidelity is even significantly improved compared to the original parameters.

\section{Conclusion\label{sec:conclusion}}

In conclusion, we have proposed a cross-resonance gate that implements an echoed conditional displacement gate that controls an oscillator with the fluxonium as the auxiliary qubit. The gate scheme includes techniques like selective darkening and echoing to ensure a high-fidelity gate. Moreover, we provide constraints and intuitions on the selection of all system and gate parameters to narrow down parameter space. Our protocol for parameter selection is numerically efficient since it only requires the spectrum and matrix elements of the fluxonium, without the need to diagonalize the whole system or to perform time domain simulations. We provide justifications to the intuitions by performing a detailed error analysis on the drive Hamiltonian, which could be categorized into on-resonance and off-resonance terms. Dominant error sources at short and long gate times have been identified. With our gate scheme, we demonstrated the ability to implement high-fidelity conditional displacements in microsecond timescales for all fluxonium parameters we survey, most of which are from experimental literatures and are available with current fabrication techniques. The best performing gate parameter we discover can drive an ECD gate with displacement length $1.6$ in $250ns$ with infidelity as low as $5\times 10^{-5}$.

While one of the key motivations of fluxonium auxiliary qubit is its biased error property, we leave the full open-system simulation to future work and restrict our focus to the gate's coherent errors and modeling. Moreover, we focused on the gate fidelity, but it could be interesting to further investigate the gate's performance in more specified applications, such as GKP stabilizations and fluxonium qubit readout. Last but not least, some gate techniques like DRAG can potentially lower the gate infidelity when the gate error is dominated by leakage at short gate times, which is a promising direction to explore.

\begin{acknowledgements}

We would like to thank Liang Jiang, Joseph Iverson, Catherine Leroux, Shraddha Singh, Shantanu Jha, and Shoumik Chowdhury for helpful discussions. We thank Simone Severini, Bill Vass, Oskar Painter, Fernando Brandão, Eric Chisholm and AWS for
supporting the quantum computing program.
\end{acknowledgements}

\appendix


\section{Derivation of the drive Hamiltonian through external flux modulation \label{sec:flux_drive_der}}
\setcounter{equation}{0}
\makeatletter
\renewcommand{\theequation}{A\arabic{equation}}

The proper placement of the time-dependent flux gives a fluxonium Hamiltonian of the form~\cite{PhysRevApplied.19.034031, PhysRevB.99.174512}
\begin{eqnarray}
    && \hat{H}_{f}(t)/h \nonumber\\
    &=& 4E_c \hat{n}^2 - E_J \cos{\hat{\varphi}} +  \frac{E_L}{2}\left(\hat{\varphi} + 2\pi \frac{\Phi_{\text{ext}}(t)}{\Phi_0}\right)^2.
\end{eqnarray}
If we apply an external flux as $\Phi_{\text{ext}}(t) = \Delta \Phi + \frac{1}{2}\Phi_0 + \frac{\Phi_0}{2\pi E_L}f(t)$, 
\rev{
\begin{eqnarray}
  && \hat{H}_{f}(t)/h \nonumber\\
    &=& 4E_c \hat{n}^2 - E_J \cos{\hat{\varphi}} \nonumber\\
    && + \frac{E_L}{2}\left(\hat{\varphi} + \pi + 2\pi \frac{\Delta \Phi}{\Phi_0} + \frac{f(t)}{E_L}\right)^2\label{eq:derive_flux_mod}\\
    &=& 4E_c \hat{n}^2 + E_J \cos{\hat{\varphi}^\prime} \nonumber\\
    && + \frac{E_L}{2}\left(\hat{\varphi}^\prime + 2\pi \frac{\Delta \Phi}{\Phi_0} + \frac{f(t)}{E_L}\right)^2.\\
    &=& 4E_c \hat{n}^2 + E_J \cos{\hat{\varphi}^\prime} \nonumber\\\
    && + \frac{E_L}{2}\left(\hat{\varphi}^\prime + 2\pi \frac{\Delta \Phi}{\Phi_0}\right)^2 + f(t)\hat{\varphi}^\prime.
\end{eqnarray}}
\rev{with $\hat{\varphi}^\prime = \hat{\varphi} + \pi$ being the flux operator with a constant offset. In the final equality, we also ignored offsets to the Hamiltonian, which only lead to global phases. }Therefore, modulating the external flux is equivalent to applying a flux drive with drive amplitude $f(t)$. Here, the external flux deivation is defined such that when $\Delta \Phi=0$, the fluxonium is exactly on the sweet spot. Therefore, setting $f(t) = A \Omega(t)\cos\omega_{\text{osc}}t$ recovers the required flux drive enevelope we defined in the main text.

\section{Derivations for the purturbative expression of displacement rates \label{sec:pert_theory_disp_rates}}
\setcounter{equation}{0}
\makeatletter
\renewcommand{\theequation}{B\arabic{equation}}

We can derive the displacement rates through simple first-order perturbation theory in the weak coupling limit, i.e. weak hybridizations. Here we consider the hybridized system with an oscillator of frequency $\omega_\text{osc}$ and a fluxonium with levels $\ket{i}$ with bare energies $E_{i}$ and transition frequencies $\Delta_{i, j} := E_{j} - E_i$. As we elaborated in the main text, the gate is activated through hybridizations between computational and non-computational states. To understand the eigenstate $\ket{\phi_{\mu, n}}$ of the hybridized system, we can proceed with a simplified model, where the fluxonium is only composed of two levels, $\ket{i}$ and $\ket{j}$. The capacitive coupling, $ig n \otimes (a - a^\dagger)$, couples bare states of the form $\ket{i}\otimes \ket{n}$ and $\ket{j}\otimes \ket{n\pm 1}$. The matrix elements are the same when the latter state is in either $\ket{n\pm 1}$. Without loss of generality, we assume $E_j\geq E_i$ and for now focus on the state with a smaller energy gap, i.e. $\ket{j}\otimes \ket{n- 1}$. The static system Hamiltonian truncated to such a subspace is
\begin{eqnarray}
    \hat{H} &=& (\Delta_{i,j} - \omega_\text{osc}) \dm{j}\otimes \dm{n-1} \nonumber\\
    && + g\sqrt{n}\sbkt{i n_{j, i} \ket{j}\bra{i}\otimes \ket{n-1}\bra{n} + \text{h.c.}}\label{eq:trunc_Ham}
\end{eqnarray}
where we defined the charge operator elements to be $n_{i,j} = \sandwich{i}{n}{j}$. For simplicity, we assume $\hbar = 1$ from here on. The hybridized state is the eigenstate, and from diagaonalization, we have 
\begin{eqnarray}
     \ket{\phi^j_{i, n}} &=& \st{i}\otimes \st{n} - \frac{ig n_{j,i}\sqrt{n}}{\Delta_{i, j} - \omega_\text{osc}}\st{j}\otimes \st{n-1} +\nonumber\\
     && \bigO{\abs{\frac{g n_{i,j}\sqrt{n}}{\Delta_{i, j} - \omega_\text{osc}}}^2}
\end{eqnarray}
where the superscript of $\ket{\phi^j_{i, n}}$ denotes the hybridization within the truncated fluxonium subspace. Such an expression can be easily extended to include contributions from both $\ket{j}\otimes \ket{n-1}$ and $\ket{j}\otimes \ket{n+1}$. As a first-order approximation based on the weak coupling constraint, $\frac{ig n_{i, j}\sqrt{n}}{\Delta_{i, j} - \omega_\text{osc}} \ll 1$ for all $i, j$, the eigenstate can be written as
\begin{eqnarray}
    && \ket{\phi_{i, n}} \approx  \st{i}\otimes \st{n} - \nonumber\\
    && \sum_{j \neq i } ig n_{j, i} \st{j}\otimes \sbkt{\frac{\sqrt{n}}{\Delta_{i,j} + \omega_\text{osc}} \st{n-1} + \frac{\sqrt{n+1}}{\Delta_{i,j} + \omega_\text{osc}} \st{n+1}}.\nonumber
\end{eqnarray}
With the expression of the eigenstates, we can now compute the matrix element of the flux drive, $\hat{\varphi}$, that generates the target dynamics. Recall that the conditional displacement Hamiltonian
\begin{eqnarray}
    \hat{H}_{\text{CD}}^\mu &\propto& \hat{U}_0 \dm{\mu}\otimes \sbkt{\hat{a} + \hat{a}^\dagger}\hat{U}_0^\dagger \nonumber\\
    &=& \sum_{n=0}^\infty \sqrt{n}\sbkt{\ket{\phi_{\mu, n}} \bra{\phi_{\mu, n-1}} + \text{h.c.}},
\end{eqnarray}
which are the matrix elements we aim to drive with the flux drive. When the systems are coupled with a coupling strength of $g_0$, we can then define the displacement rate per unit drive amplitude as
\begin{eqnarray}
    r_i &:=& g_0 \lim_{g\to 0} \frac{\bra{\phi_{i, n}} \hat{\varphi}\ket{\phi_{i, n-1}}}{g\sqrt{n}}\\
    &\approx& g_0\sum_{j \neq i}in_{i, j}\varphi_{ j,i}\sbkt{\frac{1}{\Delta_{i, j}- \omega_\text{osc}} - \frac{1}{\Delta_{i, j} + \omega_\text{osc}}}
\end{eqnarray}
where the limit is taken to get rid of higher-order terms that contribute to these matrix elements but are not proportional to $\sqrt{n}$, the bosonic enhancement factor. We have also used the fact that $in_{i, j}\varphi_{j,i}$ is real. The conditional displacement gate rate is then naturally given by 
\begin{eqnarray}
        r_{\text{CD}} := r_1 - r_0.
\end{eqnarray}

Numerically, to extract the displacement rates at finite coupling strength, we can perform the following optimization
\begin{eqnarray}
  \min_{r_i, p_i} \norm{ M }_F^2\label{eq:optimization_rates}
\end{eqnarray}
with 
\begin{eqnarray}
  && M := \nonumber\\
  &&\sum_{i, n}\sbkt{\bra{\phi_{i, n}} \hat{\varphi}\ket{\phi_{i, n-1}} - r_i \sqrt{n} - p_i n^{3/2}} \ket{\phi_{i, n}}\bra{\phi_{i, n-1}}\nonumber
\end{eqnarray}
where the third term in the definition of $M$ parameterized by $p_i$ is the rate of the leading order higher-order term. If we discard the third term, the optimized $r_i$ value will have a strong dependence on the oscillator cutoff dimension because the higher-order terms scale differently with the oscillator energy $n$. Here, the factor of $n^{3/2}$ approximately represents an ansatz of higher-order terms proportional to $\hat{a}^\dagger \hat{a} \sbkt{\hat{a}^\dagger + \hat{a}} + \text{h.c.}$. 

\begin{figure}[t!]
    \centering
    \includegraphics[width = 0.45 \textwidth]{ 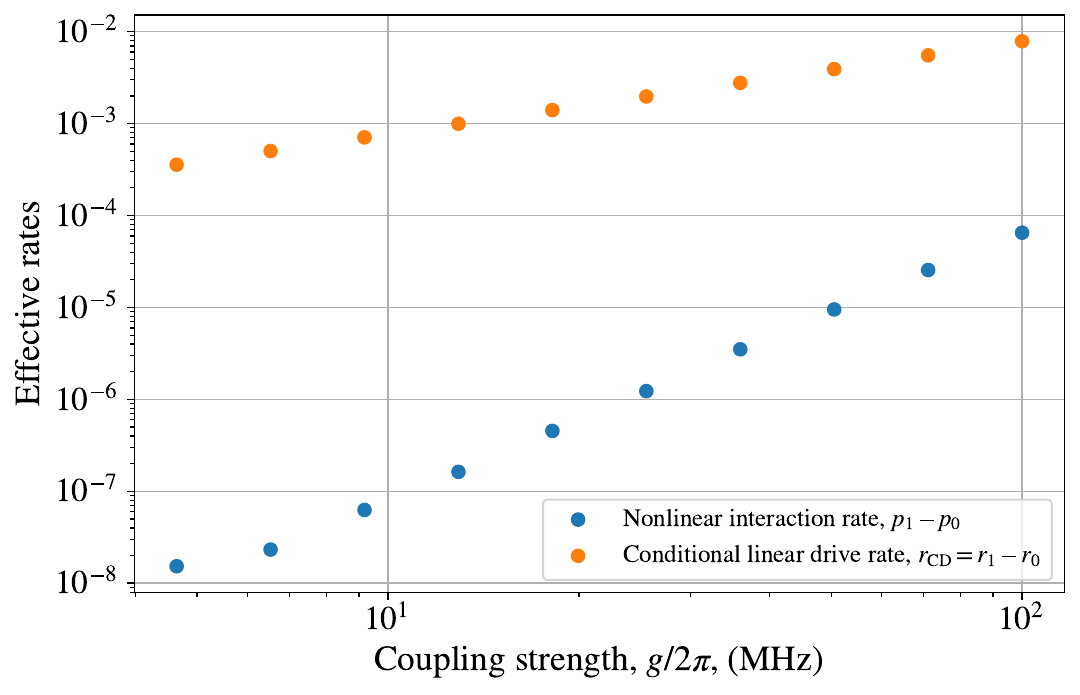}
     \caption{Conditional linear drive rate, $r_{\text{CD}}$, and the nonlinear interaction rate with varying coupling strength. The rates are obtained from numerical optimizations as formulated in Eq.~\eqref{eq:optimization_rates}. The fluxonium is chosen to be fluxonium $F_1$ in Tab.~\ref{tab:fluxonium_params}, and the oscillator frequency is $\omega_{\text{osc}}/2\pi = 5.03$ GHz.}
    \label{fig:high_order_fit}
\end{figure}

Figure.~\ref{fig:high_order_fit} demonstrates the optimization results from Eq.~\eqref{eq:optimization_rates}. The slope of the nonlinear interaction as a function of the coupling strength is indeed roughly three times that of the conditional displacement rate, which matches with our ansatz of the higher-order Hamiltonian. The optimized values of $r_{\text{CD}}$ are used in the main text to compare with the theoretical predictions from the above derivations.

\section{Gate simulation detail \label{sec:gate_details}}
\setcounter{equation}{0}
\makeatletter
\renewcommand{\theequation}{C\arabic{equation}}

As we discussed in the main text, the gate fidelity is defined as ~\cite{Pedersen_2007}
\begin{eqnarray}
    F = \frac{1}{n(n+1)}\left[\Tr{M^\dagger M} + \abs{\Tr{M}}^2\right]
\end{eqnarray}
where $\hat{P}$ is the projector onto the subspace-of-interest, $\hat{U}_t$ is the target gate, $\hat{U}$ is the actual gate, $M:= \hat{P}\hat{U}_t \hat{U} \hat{P}$, and $n = \Tr{\hat{P}}$. The target gate for our work is $\hat{U}_0 \hat{\text{ECD}}_{\alpha_0} \hat{U}_0^\dagger$ with $\alpha_0 = 1.6$. Here, we would like to point out that there are two actions that are deemed free in our simulation. The first one is the phase rotation, $\hat{U}_0 \hat{Z}_f\sbkt{\theta}\hat{U}_0^\dagger $, on the fluxonium, which can in principle be done in software. The second operation is a unconditional displacement, $\hat{U}_0 \hat{D}_\alpha\hat{U}_0^\dagger$. This displacement can be used to adjust the actual gate to match the target ECD gate. Since the displacement length is generally small, $\alpha\ll 1$, it can be realized with high fidelity if we want to physically implement it with a linear drive. Moreover,it is strictly speaking not necessary in many tasks, such as when the ECD gate is applied to readout the fluxonium state. As a result, to get an accurate evaluation of the gate fidelity, we follow the optimization below,
\begin{eqnarray}
    F &=& \max_{\theta, \alpha}  \frac{1}{n(n+1)}\left[\Tr{M^\dagger M} + \abs{\Tr{M}}^2\right]
\end{eqnarray}
where $M := \hat{P}\hat{U}_0 \hat{\text{ECD}}_{\alpha_0} \hat{Z}_f \hat{D}_\alpha\hat{U}_0^\dagger  \hat{U} \hat{P}$

\begin{figure}[t!]
    \centering
    \includegraphics[width = 0.45 \textwidth]{ 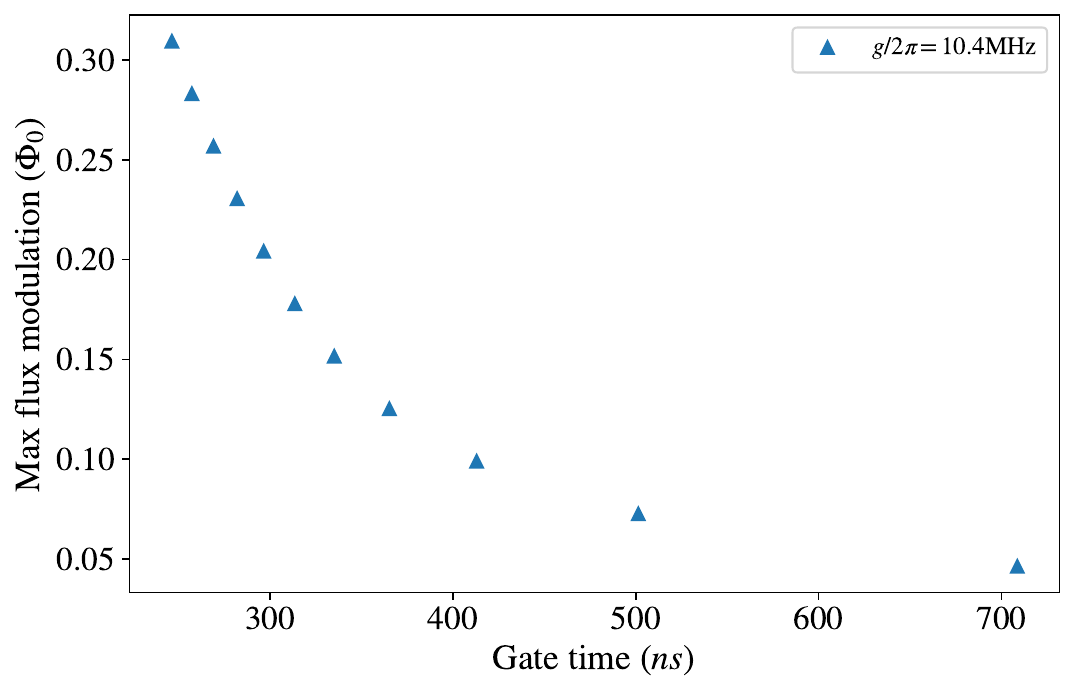}
     \caption{\rev{Maximum external flux modulation as a function of the gate time for fluxonium $F_4$ in Tab.~\ref{tab:fluxonium_params}. The max flux modulation is given in units of flux quanta, $\Phi_0$, and can be computed as $\frac{\max_{t}f(t)}{2\pi E_L}$ as shown in Eq.~\eqref{eq:derive_flux_mod}. The oscillator frequency is $\omega_{\text{osc}}/2\pi =6.0$ GHz, and the target gate is an ECD gate with displacement length $\alpha = 1.6$. }}
    \label{fig:drive_strength_nguyen}
\end{figure}

\rev{As we examine in detail in Sec.~\ref{sec:mismatch_disp_rates}, the effective displacement rate is not a constant with varying drive strength. In order to accurately implement the target gate, in simulations we determine the gate time under a fixed drive strength through an iterative approach until the target displacement length is reached. To give a sense of the drive strength required, in Fig.~\ref{fig:drive_strength_nguyen} we show some examples for displacing the oscillator with $F_4$ in Tab.~\ref{tab:fluxonium_params}. As derived in Sec.~\ref{sec:flux_drive_der}, the flux drive is realized through modulating the external flux at around the oscillator frequency. Thus, the vertical axis displays the maximum amplitude of the external flux that one needs to modulate. Here, we display the results in units of flux quanta, $\Phi_0$. Thus, the value is computed by $\frac{\max_{t}f(t)}{2\pi E_L}=\frac{A}{2\pi E_L}$, with $A$ being the flux drive amplitude and $E_L$ being the inductive energy of the fluxonium. The notation follows from derivations in Eq.~\eqref{eq:derive_flux_mod}.}

\rev{The system parameters are identical to those adopted in Fig.~\ref{fig:survey} (d). Here, for the shortest gate time, the maximum external flux modulation is roughly $0.3\Phi_0$. To put the number into context, tuning the external flux over half flux quantum, $0.5\Phi_0$, covers the minimum and maximum fluxonium frequency achievable by controlling the external flux. As we modulate the external flux, the eigenbasis and spectrum of the fluxonium could vary significantly, so it is crucial to ensure that a sufficient size of Hilbert space is included in the simulations. Meanwhile, it is remarkable that the operating regime obtained through our non-perturbative analysis is still applicable at varying gate regimes. In practice, driving the system too hard could also lead to additional noise from, for example, stray fields. The exact trade-off could vary from one hardware to another. Moreover, we remark that the gate times we demonstrate are relatively short compared to other existing CD gates. Therefore, the gate can operate with weaker drives if necessary depending on the experimental conditions.
}

\section{Derivations for the mismatched displacement rates \label{sec:mismatch_disp_rates}}
\setcounter{equation}{0}
\makeatletter
\renewcommand{\theequation}{D\arabic{equation}}

\begin{figure*}
  \centering
  \includegraphics[width = 0.9 \textwidth]{ 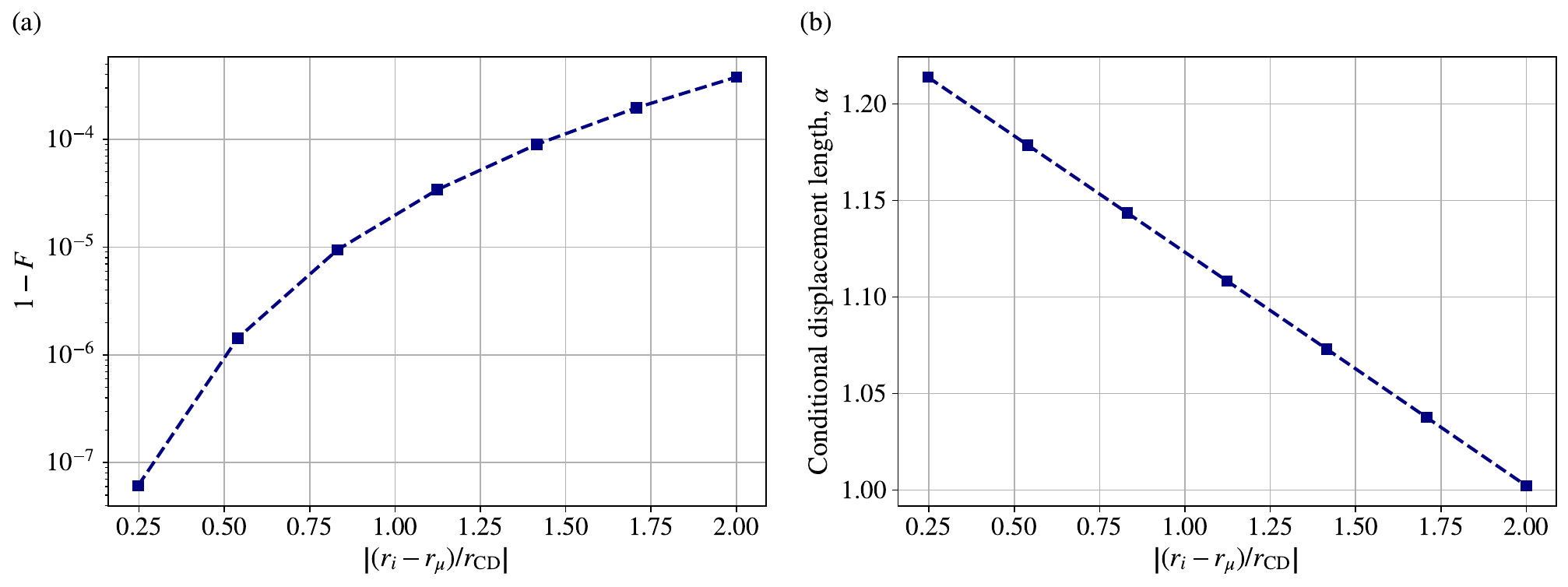}
   \caption{The (a) infidelity and (b) CD gate displacement length of the simplified model with varying ratios between the displacement rates, $\abs{(r_i - r_\mu)/r_{\text{CD}}}$. We fix the drive strength $A/2\pi = 0.95$GHz, gate time $T_{g} = 380$ns, $r_{\text{CD}} = 2.4\times 10^{-3}$, and the pulse form to be the same as that given in Eq.~\eqref{eq:waveform}.}
  \label{fig:toy_model_mism_disp_rate}
\end{figure*}

This section derives the model and effect of the mismatched displacement rates. First, we follow the same assumption in the main text, where we constrain that only one transition, $\sbkt{\mu, i}$, gives a dominant contribution to the gate rate. Suppose we approximate the fluxonium as a two-level system, $\set{\ket{\mu}, \ket{i}}$, and redefine the Pauli operators within this subspace as $\hat{\sigma}^\prime_k$ with $k = x, y, z$. As shown in Appendix~\ref{sec:pert_theory_disp_rates}, the capacitive coupling give rise to the hybridization between states of the form $\ket{\mu}\otimes \ket{n}$ and $\ket{i}\ket{n\pm 1}$, which lead to conditional displacement rates to first order in the coupling strength. Under the assumption of weak drive and only keeping the induced dynamics to first order, we can approximate the system as being uncoupled, but with a flux drive that effectively drives conditional displacements. The truncated full system Hamiltonian is
\begin{eqnarray}
                && \hat{H} \approx\frac{\Delta_{\mu, i}}{2}\hat{\sigma}^\prime_z + \omega_\text{osc} \hat{a}^\dagger \hat{a} + A \Omega(t) \cos\omega_\text{osc} t\times\nonumber\\
                && \{\varphi_{\mu, i}^x\hat{\sigma}^\prime_x + \varphi_{\mu, i}^z\hat{\sigma}^\prime_z + \sbkt{r_{i} \dm{i} + r_{\mu} \dm{\mu}} \otimes (\hat{a} + \hat{a}^\dagger)\}\nonumber
\end{eqnarray}
where we defined the flux operator projected within the subspace to be $\varphi_{\mu, i}^x := \bra{i}\hat{\varphi}\ket{\mu}$ and $\varphi_{\mu, i}^z := \frac{1}{2}\sbkt{\bra{i}\hat{\varphi}\ket{i} - \bra{\mu}\hat{\varphi}\ket{\mu}}$. Here, we note that the flux operator do not have $\hat{\sigma}_{\mu, i}^y$ components in general. In the rotating frame $e^{i\omega_{\text{osc}}\left(\frac{\hat{\sigma}^\prime_z}{2} + \hat{a}^\dagger \hat{a}\right) t}$, we can make the rotating wave approximation and obtain
\begin{eqnarray} 
    && \hat{H}_I = \frac{\Delta_{\mu, i} - \omega_{\text{osc}}}{2}\hat{\sigma}^\prime_z + \frac{A\Omega(t)}{2} \times \nonumber\\
    &&\sbkt{\varphi_{\mu, i}^x\hat{\sigma}^\prime_x + \sbkt{r_{i} \dm{i} + r_{\mu} \dm{\mu}} \otimes (\hat{a} + \hat{a}^\dagger)}. \nonumber
\end{eqnarray}
The pulse envelope is ideally a smooth function that varies adiabatically and $\Omega(0) = \Omega(T_{g}) = 0$, with $T_g$ being the gate time. Such adiabaticity reduces leakage if, for example, the fluxonium state is initialized in $\ket{\mu}$. Ignoring the third term, we can derive the time-dependent eigenstates of the first two terms of $\hat{H}_I$ to first order in drive strength as 
\begin{eqnarray}
    \vert \tilde{i}\rangle   &\approx& \st{i} + \frac{A\Omega(t) \varphi_{\mu, i}^x}{\Delta_{\mu, i} - \omega_{\text{osc}}} \st{\mu}\\
    \st{\tilde{\mu}}  &\approx& \st{\mu} - \frac{A\Omega(t) \varphi_{\mu, i}^x}{\Delta_{\mu, i} - \omega_{\text{osc}}} \st{i},
\end{eqnarray}
and they correspond to energies $E_i(t)$, $E_\mu (t)$ respectively. Under weak drive limit, $\epsilon:= \frac{A\Omega(t) \varphi_{\mu, i}^x}{\Delta_{\mu, i} - \omega_{\text{osc}}}\ll 1$. To leading order, the Hamiltonian can be rewritten as 
\begin{eqnarray}
                 && H_I \approx E_\mu(t) \dm{\tilde{\mu}} + E_i(t) \vert \tilde{i}\rangle\langle \tilde{i}\vert +\frac{A\Omega(t)}{2}\times \nonumber\\
                 && \mbkt{\sbkt{r_\mu + (r_i - r_\mu) \epsilon^2}\dm{\tilde{\mu}}  + \sbkt{r_i + (r_\mu - r_i)  \epsilon^2}\vert \tilde{i}\rangle\langle \tilde{i}\vert}\otimes \nonumber\\
                 && \sbkt{a + a^\dagger}+ \frac{A\Omega(t)}{2} \epsilon \sbkt{r_\mu - r_i}\sigma^\prime_x(t) \otimes \sbkt{a + a^\dagger}\nonumber
\end{eqnarray}
where the time-dependent Pauli operator is $\sigma^\prime_x(t) = \vert \tilde{i}\rangle\bra{\tilde{\mu}} + \text{h.c.}$. The second line represents the effective conditional drives. Notice that in the adiabatic limit and ignore imperfections brought by the third line, the fluxonium with initial state $\ket{\mu}$ is expected to stay in $\ket{\tilde{\mu}}$. Therefore, the effective displacement rate is time-dependent with the form of
\begin{eqnarray}
    r_{\mu}(t) = r_\mu + (r_i - r_\mu)\abs{\frac{A\Omega(t)}{\Delta_{\mu, i} - \omega_{\text{osc}}}\bra{i}\hat{\varphi}\ket{\mu}}^2.
\end{eqnarray}
The last line represents a Hamiltonian that drives the oscillator while inducing fluxonium transitions among $(\mu, i)$. Such a term deviates from the target condition displacement gate and introduces error, and it is proportional to $\sbkt{A\Omega(t)}^2\frac{ \varphi_{\mu, i}^x \sbkt{r_\mu - r_i}}{2\sbkt{\Delta_{\mu, i} - \omega_{\text{osc}}}}$. Intuitively, this is reasonable since if $r_\mu = r_i$, the oscillator dynamics is identical regardless of the fluxonium state. Therefore, to suppress the induced error, it is preferable to reduce the mismatched displacement rate, $\abs{r_\mu - r_i}$, and operate at weak drive strength.

To give an example, we simulate the toy model with a 3 level system and an oscillator. All gate and system parameters are taken to be the same as a CD gate implemented with fluxonium $F_1$. In particular, we fix the drive strength $A/2\pi = 0.95$GHz, gate time $T_{g} = 380$ns, $r_{\text{CD}} = 2.4\times 10^{-3}$, and the same pulse form as given in Eq.~\eqref{eq:waveform}. The ratio of the mismatched displacement rate is varied. It is clear that the compromise in the gate displacement rate is linearly proportional to the ratio. Moreover, the CD gate infidelity increases with increasing ratio. However, as we mentioned in the main text, this simplified model cannot by itself explain the dominant error source at such gate times. For example, while the ratio of the mismatched displacement rates is in only $\sim 5\%$ for the cavity frequency we consider, the infidelity of the full simulation of the gate with fluxonium $F_1$ is on the level of $10^{-4}$ for this particular gate time and drive strength we consider.

\section{\rev{Supplemental simulation of gate performance}}
\setcounter{equation}{0}
\makeatletter
\renewcommand{\theequation}{E\arabic{equation}}

\begin{figure}[t!]
    \centering
    \includegraphics[width = 0.45 \textwidth]{ 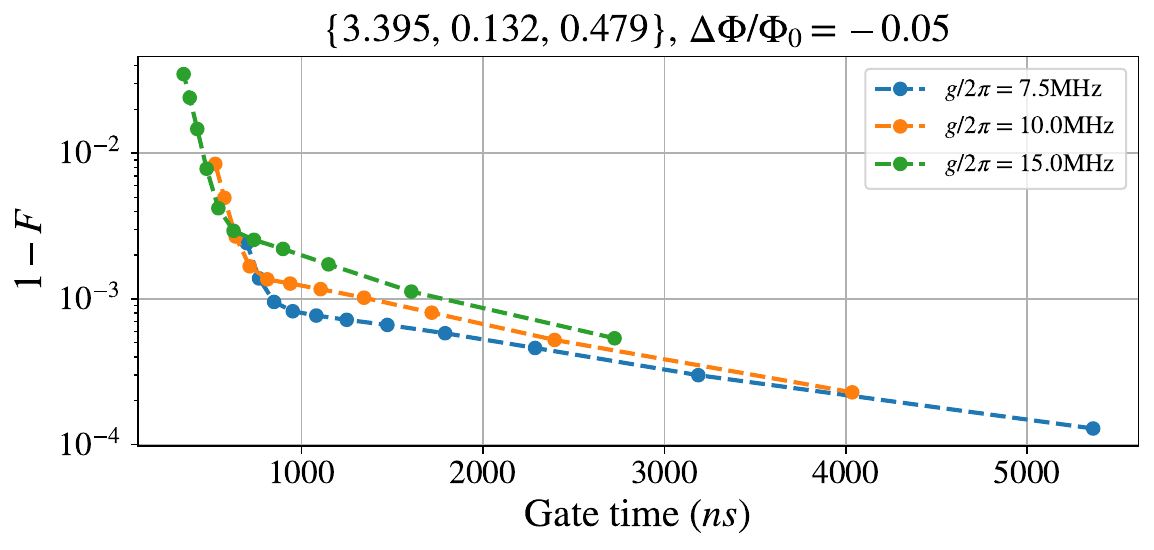}
     \caption{\rev{Gate performance of a heavy fluxonium operated off sweet spot. The fluxonium energy parameter sets are labelled in the order of $\left\{E_J/h, E_L/h, E_C/h\right\}$, all in units of GHz. The oscillator frequency is $\omega_{\text{osc}}/2\pi = 5.2$ GHz. The target gate is an ECD gate with displacement length $\alpha = 1.6$.}}
    \label{fig:schuster_offss_performance}
\end{figure}
\rev{
In Sec.~\ref{sec:survey}, we have presented a survery of numerous fluxonium parameters and their performance under the proposed gate scheme. The examples presented include both on sweet spot and off sweet spot fluxoniums. Here, we include some additional simulations performed for fluxonium parameters $\left\{E_J/h, E_L/h, E_C/h\right\} = \left\{3.395, 0.132, 0.479\right\}$ with all units in GHz. This set of energy parameters corresponds to $F_1$ in Table.~\ref{tab:fluxonium_params}. Nevertheless, we operate the fluxonium at a external flux of $\Delta \Phi/\Phi_0 = -0.05$, which is different from the main text. The simulation results are shown in Fig.~\ref{fig:schuster_offss_performance}, and it is clear that the gate's unitary fidelity is decent at relatively longer gate times. }

\rev{We include the results for this specific fluxonium because it is a heavy fluxonium operating off sweet spot regime. This regime is where where one would expect a significantly suppressed bit flip rate. For example, for this specific fluxonium, the $T_1$ time is increased from around $200\mu$s to around $4$ms by going off sweet spot~\cite{Zhang_2021}. 
}

\end{document}